\documentclass[11pt]{article}

\PassOptionsToPackage{nosort}{cite}

\usepackage{tikz}
\usepackage{float}
\usepackage{mathtools}
\usepackage{scalerel}
\usepackage{amssymb}
\usepackage{amsfonts}
\usepackage{amsthm}
\usepackage{dsfont}
\usepackage{mathrsfs}
\usepackage{enumitem}
\usepackage[T1]{fontenc}
\usepackage{chet}
\usepackage{slashed}
\usepackage{multirow}
\usepackage[font=small,format=hang,labelfont={sf,bf}]{caption}
\usepackage{subcaption}

\usetikzlibrary{intersections,arrows.meta}

\definecolor{darkblue}{rgb}{0.1,0.1,0.7}
\hypersetup{colorlinks,
           linkcolor={darkblue},
           citecolor={darkblue},
           urlcolor={darkblue},
           pdftitle={Line Defect RG Flows in the Epsilon Expansion},
           pdfdisplaydoctitle
}

\tikzset{cross/.style={path picture={
      \draw[black]
            (path picture bounding box.south east) --
            (path picture bounding box.north west)
            (path picture bounding box.south west) --
            (path picture bounding box.north east);}}}

\newcommand{\lsp}{\hspace{0.5pt}}

\renewcommand{\geq}{\geqslant}
\renewcommand{\leq}{\leqslant}
\newcommand{\overbar}[1]{\mkern 1.5mu\overline{\mkern-1.5mu#1\mkern-1.5mu}\mkern 1.5mu}

\DeclareMathOperator{\Tr}{Tr}

\title{Line Defect RG Flows in the $\varepsilon$ Expansion}

\author{William H.\ Pannell and Andreas Stergiou\emails{(\href{mailto:william.pannell@kcl.ac.uk}{william.pannell}, \href{mailto:andreas.stergiou@kcl.ac.uk}{andreas.stergiou})@kcl.ac.uk}}

\affiliation{Department of Mathematics, King's College London, Strand, London WC2R 2LS, United Kingdom}

\abstract{A general analysis of line defect renormalisation group (RG) flows in the $\varepsilon$ expansion below $d=4$ dimensions is undertaken. The defect beta function for general scalar-fermion bulk theories is computed to next-to-leading order in the bulk couplings. Scalar models as well as scalar-fermion models with various global symmetries in the bulk are considered at leading non-trivial order. Different types of potential infrared (IR) defect conformal field theories (dCFTs) and their RG stability are discussed. The possibility of multiple IR stable dCFTs is realised in specific examples with hypertetrahedral symmetry in the bulk. The one-point function coefficient of the order parameter in the stable IR dCFT of the cubic model is computed at next-to-leading order and compared with that in the IR dCFT of the Heisenberg model.}

\date{March 2023}

\begin{document}

\maketitle

\toc

\section{Introduction}
The $\varepsilon$ expansion appeared more than 50 years ago~\cite{Wilson:1971dc} in an attempt to access the physics of critical models in $d=3$ dimensions in a perturbative way. To this day it remains one of the most powerful and versatile tools at our disposal for the study of the renormalisation group (RG) and conformal field theories (CFTs). The most widely employed strategy has been to impose different global symmetries and seek RG fixed points that may describe critical points of corresponding physical systems. In this work we discuss the behaviour of such fixed points when deformed by line defect operators.

In a $d$-dimensional bulk CFT the introduction of a straight line defect operator will result in the breaking of the conformal group $SO(d+1,1)$. If a fixed point of the system with the insertion of the line defect operator exists, then the spacetime symmetry group preserved at this fixed point of the combined bulk-defect system, i.e.\ at the defect CFT (dCFT), is $SL(2,\mathbb{R})\times SO(d-1)$. Any global symmetry of the bulk CFT will generically also be broken to a subgroup in the dCFT.

Concretely, the line defect deformations we consider in this work take the form
\begin{equation}\label{defdeform}
S_{\text{defect}}(h,\phi)=h_i\int_{-\infty}^{
\infty} d\tau\, \phi_i(\tau,\mathbf{0})\,,\qquad i=1,\ldots,N,
\end{equation}
where $\phi_i$ are scalar fields with dimension below 1 so that the deformation is relevant and $h_i$ are defect couplings. One may think of $h_i$ as a background magnetic (pinning) field coupled to the order parameter $\phi_i$. These deformations are added to bulk CFTs with various global symmetry groups $G$. We generically find that there are non-trivial critical values of $h_i$ for which the defect RG flow terminates at a critical point in the infrared (IR), where the associated beta functions, $\beta_i=\mu\lsp dh_i/d\mu$, become zero. Such vectors $h$ break $G$ to a proper subgroup $K<G$ on the defect. The beta function is $G$-covariant, and so $gh$, $g\in G$, will define a dCFT if $h$ does.\footnote{Eq.\ \eqref{defdeform} can be used to define the action of $G$ on the space of $h$ vectors from its action on the space of $\phi$ vectors.} If $g\notin K$, then the vectors $h$ and $gh$ will be different. Nevertheless, dCFTs corresponding to vectors $h$ related by an action of $G$ are equivalent. There is a one-to-one correspondence between such vectors and the cosets of $K$ in $G$, and so the set of such vectors is isomorphic to the quotient $G/K$. Defect CFTs in $G/K$ evoke the situation of degenerate vacua in the case of spontaneous symmetry breaking. In the case of continuous $G$ the broken currents of the bulk CFT give rise to marginal operators in the dCFT, whose correlation functions can be used to describe the geometry of the manifold $G/K$~\cite{Drukker:2022pxk}.

The defect breaks translation invariance in the transverse directions, which implies that one-point functions in the bulk can be non-zero. This turns out to be useful in numerical simulations of critical theories~\cite{Assaad:2013xua, ParisenToldin:2016szc}. In statistical systems, one can find spontaneous symmetry breaking by studying the behaviour of the relevant order parameter. Near critical points, this usually necessitates the calculation of the local order parameter's two-point function, which is quadratically suppressed in cases where the order parameter is small. This can cause difficulties, especially in numerical simulations of these systems where high accuracy will be required to compensate. As a remedy to this problem, one can consider introducing a defect into the system. Though the system's original symmetry will now be explicitly broken, far from the defect the bulk parameters will be unaffected, thus acting as though the symmetry remained. The critical behaviour of the system can then be investigated by simply looking at the one-point function of the order parameter.

Studies of line defects in the $\varepsilon$ expansion have so far focused mostly on the scalar $O(N)$ model~\cite{Allais:2014fqa, Cuomo:2021kfm}. Here the defect deformation is added to the bulk $O(N)$-symmetric theory at its non-trivial RG fixed point. The deformation is relevant and thus a defect RG flow ensues, which ends up terminating at a non-trivial dCFT in the IR, i.e.\ one in which the associated defect coupling is non-zero. Throughout this defect RG flow the bulk theory remains at its fixed point. Further studies in the same framework include line defects in tensor models with $O(N)^3$ symmetry~\cite{Popov:2022nfq}, and the Gross--Neveu--Yukawa model, which involves scalars and fermions~\cite{Giombi:2022vnz}. On the contrary,~\cite{Rodriguez-Gomez:2022gbz, Rodriguez-Gomez:2022gif} considered a double-scaling limit that renders the bulk theory classical and analysed defect deformations without imposing that the bulk theory be critical.

There exists a variety of bulk critical models one can obtain in the $\varepsilon$ expansion; see~\cite{Pelissetto:2000ek, Osborn:2017ucf, Rychkov:2018vya, Osborn:2020cnf} for such models in $d=4-\varepsilon$ with scalar fields only. Many of them possess only one quadratic invariant, namely the $\phi^2$ operator, but there exist cases with more than one such operator, commonly referred to as biconical or, more generally, multiconical. In this work we discuss the $O(N)$, hypercubic, hypertetrahedral and MN models, and also the $O(m)\times O(n)$ biconical model. There are also a number of bulk theories one can obtain by introducing fermions, and there has been some recent interest in exploring these models, especially those which give rise to emergent supersymmetry in three dimensions\cite{Jack:2023zjt, Liendo:2021wpo, Fei:2016sgs}. Here, we consider the Gross--Neveu--Yukawa, Nambu--Jona-Lasinio--Yukawa and chiral Heisenberg models. In all these models there exist $N$ scalar fields $\phi_i$ with dimension below 1, and the deformation \eqref{defdeform} triggers an RG flow that generically terminates at an IR fixed point with real critical values for $h_i$.

One might expect that among the possible IR dCFTs there will exist only one without relevant operators, so that the RG flow from the UV will generically terminate there. In the $\varepsilon=4-d$ expansion for scalar CFTs without defect deformations, there exists a theorem that shows that if an RG stable CFT exists in the IR of a given set of relevant perturbations of a UV CFT, then it is unique~\cite{Michel:1983in}. We find that this is not necessarily the case for scalar dCFTs, by exhibiting specific examples of multiple stable IR dCFTs with hypertetrahedral global symmetry in the bulk. These dCFTs have different global symmetries and disjoint basins of attraction so that there are no RG flows connecting them.

Our discussion of dCFTs is divided in the following manner: In section \ref{betasec} we derive the beta function for the defect couplings $h_i$ in a theory involving a general scalar quartic interaction by examining the scalar one-point function. We then present the beta function when one adds fermions to the bulk theory, postponing until the appendix a detailed derivation of the additional defect counterterms, and then examine the dependence of a subset of counterterms on bulk wavefunction renormalization. These calculations are performed to next-to-leading order in the bulk couplings. In section \ref{scalarsec} we use the general form of the defect beta function to analyse a series of example scalar theories in the bulk, using results for the fixed points of these theories to derive both the defect fixed points and examine their stability properties. Section \ref{fermionsec} then provides a similar discussion for a number of bulk scalar-fermion theories. We conclude in section \ref{conc} with a discussion of one-point functions of the order parameter in the Heisenberg and cubic models in the presence of a line defect.

\section{Defect beta function}\label{betasec}
In this section we describe in detail the computation of the beta function of the line defect coupling for theories with scalars and fermions in the bulk. Assuming a bulk CFT with scalars $\phi_i$ of dimension below 1, we will consider the relevant line-defect deformation \eqref{defdeform}. We work at next to leading order in the bulk couplings but non-perturbatively in the defect coupling. Repeated indices are assumed to be summed over the values they can take, unless otherwise indicated.

\subsection{Only scalars in the bulk}
Starting with only scalars in the bulk, the beta function has already been reported to next-to-leading order in the bulk coupling in~\cite{Allais:2014fqa, Cuomo:2021kfm} for the $O(N)$ model, where it was obtained by requiring a finite one-point function $\langle\phi_i(x)\rangle$ in the presence of the defect. Here we will use the same logic but will be more general and start with the bulk action
\begin{equation}\label{laggen}
    S=\int d^{\lsp d}x\,\big(\tfrac12\partial^\mu\phi_i\lsp\partial_\mu\phi_i+\tfrac{1}{24}\lambda_{ijkl}\phi_i\phi_j\phi_k\phi_l\big)\,,\qquad i=1,\ldots,N\,,\quad d=4-\varepsilon\,,
\end{equation}
which describes a variety of critical bulk models; see e.g.~\cite{Pelissetto:2000ek, Osborn:2017ucf, Rychkov:2018vya}. To this we will add the deformation \eqref{defdeform}. We will perform our computation of the beta function of $h_i$ in the standard paradigm of renormalised perturbation theory.\footnote{We are always free to make an $O(N)$ field rotation as long as we also rotate the couplings. Thus, we have the freedom to fix the form of either $\lambda_{ijkl}$ or $h_i$. We could choose to only look for solutions which have a single non-zero defect coupling, say $h_1$. However, without already knowing the defect fixed points for a given bulk system, one would not know how this rotation would affect the quartic coupling $\lambda_{ijkl}$. It seems much simpler to instead work with a fixed $\lambda_{ijkl}$ and a general $h_i$.} For the calculation, we will group diagrams by the order of the bulk quartic couplings and compute divergences using dimensional regularisation within the minimal subtraction (MS) renormalisation scheme.

\begin{figure}[ht]
    \centering
    \begin{tikzpicture}[baseline=(vert_cent.base)]
        \node (vert_cent) at (0,1) {$\phantom{\cdot}$};
        \draw[dashed] (0,2)--(0,0);
        \draw[thick] (-1.25,0)--(1.25,0);
        \filldraw (0,0) circle (2pt);
    \end{tikzpicture}\quad
    \begin{tikzpicture}[baseline=(vert_cent.base)]
        \node (vert_cent) at (0,1) {$\phantom{\cdot}$};
        \draw[dashed] (0,2)--(0,0);
        \draw[thick] (-1.25,0)--(1.25,0);
        \filldraw (0,0) circle (2pt);
        \node [fill, shape=rectangle, minimum width=4pt, minimum height=4pt, inner sep=0pt, anchor=center] at (0,1) {};
        \draw[dashed] (0,1)--(-0.625,0);
        \filldraw (-0.625,0) circle (2pt);
        \draw[dashed] (0,1)--(0.625,0);
        \filldraw (0.625,0) circle (2pt);
    \end{tikzpicture}\quad
    \begin{tikzpicture}[baseline=(vert_cent.base)]
        \node (vert_cent) at (0,1) {$\phantom{\cdot}$};
        \draw[dashed] (0,2)--(0,0);
        \draw[thick] (-1.25,0)--(1.25,0);
        \filldraw (0,0) circle (2pt);
        \node [fill, shape=rectangle, minimum width=4pt, minimum height=4pt, inner sep=0pt, anchor=center] at (0,1.5) {};
        \node [fill, shape=rectangle, minimum width=4pt, minimum height=4pt, inner sep=0pt, anchor=center] at (0,0.5) {};
        \draw[dashed] (0,1) circle (0.5cm);
    \end{tikzpicture}\quad
    \begin{tikzpicture}[baseline=(vert_cent.base)]
        \node (vert_cent) at (0,1) {$\phantom{\cdot}$};
        \draw[dashed] (0,2)--(0,1.5);
        \draw[dashed] (0,1.5)--(1,0);
        \draw[thick] (-1.25,0)--(1.25,0);
        \node [fill, shape=rectangle, minimum width=4pt, minimum height=4pt, inner sep=0pt, anchor=center] at (-0.5,0.75) {};
        \draw[dashed] (-0.5,0.75)--(0,0);
        \draw[dashed] (-0.5,0.75)--(-1,0);
        \node [fill, shape=rectangle, minimum width=4pt, minimum height=4pt, inner sep=0pt, anchor=center] at (0,1.5) {};
        \filldraw (0,0) circle (2pt);
        \filldraw (-1,0) circle (2pt);
        \filldraw (1,0) circle (2pt);
        \draw[dashed] (0,1.5) to[out=180,in=120] (-0.5,0.75);
        \draw[dashed] (0,1.5) to[out=285,in=0] (-0.5,0.75);
    \end{tikzpicture}\quad
    \begin{tikzpicture}[baseline=(vert_cent.base)]
        \node (vert_cent) at (0,1) {$\phantom{\cdot}$};
        \draw[dashed] (0,2)--(0,0);
        \draw[thick] (-1.25,0)--(1.25,0);
        \filldraw (0,0) circle (2pt);
        \node [fill, shape=rectangle, minimum width=4pt, minimum height=4pt, inner sep=0pt, anchor=center] at (0,0.7) {};
        \draw[dashed] (0,0.7)--(-0.5,0);
        \filldraw (-0.5,0) circle (2pt);
        \draw[dashed] (0,0.7)--(0.5,0);
        \filldraw (0.5,0) circle (2pt);
        \node [fill, shape=rectangle, minimum width=4pt, minimum height=4pt, inner sep=0pt, anchor=center] at (0,1.4) {};
        \draw[dashed] (0,1.4)--(-1,0);
        \filldraw (-1,0) circle (2pt);
        \draw[dashed] (0,1.4)--(1,0);
        \filldraw (1,0) circle (2pt);
    \end{tikzpicture}
    \caption{Diagrams that contribute to $\langle\phi_i(x)\rangle$ up to next-to-leading order in the bulk quartic coupling. Squares and circles denote bulk and defect couplings, respectively. The defect is represented by the solid horizontal line.}
    \label{BetaDefDiagScalar2Loops}
\end{figure}
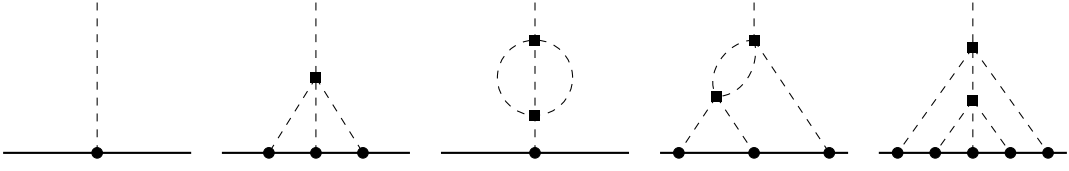

The relevant diagrams up to quadratic order in the bulk quartic coupling are shown in Fig.\ \ref{BetaDefDiagScalar2Loops}. For the calculation of these diagrams one uses the following rules:
\begin{equation}
\begin{aligned}
    \begin{tikzpicture}[baseline=(vert_cent.base)]
        \node (vert_cent) at (0,0) {$\phantom{\cdot}$};
        \draw[dashed] node[at start,xshift=-7pt,yshift=-1pt] {$x_1$} (0,0)--(1,0) node[at end,xshift=7pt,yshift=-1pt] {$x_2$};
    \end{tikzpicture} &= \frac{\Gamma(\tfrac12d-1)}{4\pi^{d/2}}\frac{1}{(x_{12}^{\;\;\; 2})^{\frac12d-1}}\,,\qquad x_{12}=x_1-x_2\,,\\
    \begin{tikzpicture}[baseline=(vert_cent.base)]
        \node (vert_cent) at (0,0) {$\phantom{\cdot}$};
        \draw[dashed,shorten <=0.75pt] (-0.5,-0.5)--(0.5,0.5);
        \draw[dashed,shorten <=0.75pt]  (-0.5,0.5)--(0.5,-0.5);
        \node[xshift=-3pt,yshift=3pt] at (-0.5,0.5) {$i$};
        \node[xshift=3pt,yshift=3pt] at (0.5,0.5) {$j$};
        \node[xshift=3pt,yshift=-3pt] at (0.5,-0.5) {$k$};
        \node[xshift=-3pt,yshift=-3pt] at (-0.5,-0.5) {$l$};
        \node [fill, shape=rectangle, minimum width=4pt, minimum height=4pt, inner sep=0pt, anchor=center] at (0,0) {};
        \node[xshift=7pt] at (0,0) {$x$};
    \end{tikzpicture} &= -\mu^{\varepsilon}\lambda_{ijkl}\int d^{\lsp d}x\,,\qquad\qquad
    \begin{tikzpicture}[baseline=(vert_cent.base)]
        \node (vert_cent) at (0,0.5) {$\phantom{\cdot}$};
        \draw[dashed]  (0,1)--(0,0) node[at start] {$i$};
        \draw[thick] (-0.5,0)--(0.5,0);
        \filldraw (0,0) circle (2pt) node[below=1pt] {$\tau$};
    \end{tikzpicture} = -\mu^{\varepsilon/2}h_i\int_{-\infty}^\infty d\tau\,.
\end{aligned}
\label{scalarrules}
\end{equation}
Here we have introduced a scale $\mu$ (of mass dimension equal to 1) so that we work with dimensionless couplings. Note that what we are computing below is $\mu^{\varepsilon/2}\langle\phi_i(x)\rangle$, which has dimension 1 classically.

For the first diagram of Fig.~\ref{BetaDefDiagScalar2Loops} we find the finite in the $\varepsilon\to0$ limit result
\begin{equation}
    \begin{aligned}
        \begin{tikzpicture}[baseline=(vert_cent.base)]
            \node (vert_cent) at (0,0.5) {$\phantom{\cdot}$};
            \draw[dashed] (0,1)--(0,0) node[at start,yshift=7pt] {$x,i$};
            \draw[thick] (-0.75,0)--(0.75,0);
            \filldraw (0,0) circle (2pt) node[below] {$\tau'$};
        \end{tikzpicture}&=-\mu^{\varepsilon}h_i\int d\tau'\,\frac{\Gamma(\tfrac12d-1)}{4\pi^{\frac12d}}\frac{1}{\big((\tau-\tau')^2+\mathbf{x}^2\big)^{\frac12(d-2)}}\\[-12pt]
        &=-\frac{\Gamma\big(\frac12(d-3)\big)}{4\pi^{\frac12(d-1)}}\frac{\mu^{\varepsilon}}{|\mathbf{x}|^{d-3}}h_i\\
        &=-\frac{1}{4\pi|\mathbf{x}|}h_i\big(1+\tfrac12(\gamma+\log(4\pi|\mathbf{x}|^2\mu^2))\varepsilon+\text{O}(\varepsilon^2)\big)\,,
    \end{aligned}
    \label{eqfinite}
\end{equation}
where $\gamma\approx0.577216$ is Euler's constant. Notice that this result is $\tau$-independent, where $x=(\tau,\mathbf{x})$. Since we will be integrating over the defect insertions, we can always substitute propagators that end on the defect with $-\frac{\Gamma(\frac12(d-3))}{4\pi^{\frac12(d-1)}}\frac{1}{|\mathbf{x}|^{d-3}}h_i$, where $\mathbf{x}$ is the spatial part of the location in the bulk from which the propagator originates.

For the second diagram of Fig.~\ref{BetaDefDiagScalar2Loops} we have
\begin{equation}\label{diag2comp}
\begin{aligned}
    \begin{tikzpicture}[baseline=(vert_cent.base)]
        \node (vert_cent) at (0,1) {$\phantom{\cdot}$};
        \draw[dashed] (0,2)--(0,0) node[at start,yshift=7pt] {$x,i$};
        \draw[thick] (-1.25,0)--(1.25,0);
        \filldraw (0,0) circle (2pt) node[below] {$\tau_2'$};
        \node [fill, shape=rectangle, minimum width=4pt, minimum height=4pt, inner sep=0pt, anchor=center] at (0,1) {};
        \node[xshift=8.5pt,yshift=-1pt] at (0,1) {$x_1$};
        \draw[dashed] (0,1)--(-0.625,0) node[midway,xshift=-6pt] {$j$};
        \filldraw (-0.625,0) circle (2pt) node[below] {$\tau_1'$};
        \draw[dashed] (0,1)--(0.625,0) node[midway,xshift=7pt] {$l$};
        \filldraw (0.625,0) circle (2pt) node[below] {$\tau_3'$};
        \node[xshift=4.5pt] at (0,0.35) {$k$};
    \end{tikzpicture}&=\tfrac16\mu^{3\varepsilon}\lambda_{ijkl}h_jh_kh_l\left(\frac{\Gamma(\frac12(d-3))}{4\pi^{\frac12(d-1)}}\right)^3\frac{\Gamma(\frac12d-1)}{4\pi^{\frac12d-1}}\int d^{\lsp d}x_1\,\frac{1}{((x-x_1)^2)^{\frac12d-1}}\frac{1}{|\mathbf{x}_1|^{3(d-3)}}\\[-1cm]
    &=\tfrac16\mu^{3\varepsilon}\lambda_{ijkl}h_jh_kh_l\left(\frac{\Gamma(\frac12(d-3))}{4\pi^{\frac12(d-1)}}\right)^4\int d^{\lsp d-1}\mathbf{x}_1\,\frac{1}{|\mathbf{x}_1-\mathbf{x}|^{d-3}|\mathbf{x}_1|^{3(d-3)}}\\
    &=-\lambda_{ijkl}h_jh_kh_l\frac{\Gamma(\frac12(d-3))^3}{768\pi^{\frac32(d-1)}(d-4)(3d-11)}\frac{\mu^{3\varepsilon}}{|\mathbf{x}|^{3d-11}}\\
    &=\frac{1}{64\pi^3|\mathbf{x}|}\Big(\frac{1}{12\lsp\varepsilon}+\frac18\big(2+\gamma+\log(4\pi|\mathbf{x}|^2\mu^{2})\big)+\text{O}(\varepsilon)\Big)\lambda_{ijkl}h_jh_kh_l\,.
\end{aligned}
\end{equation}
As we have already mentioned, we will be working in the MS scheme, but $\overline{\text{MS}}$ can be also used by sending $\mu^2\to \mu^2 e^{-\gamma}/4\pi$ so that factors of $\gamma$ or $\log 4\pi$ are removed. To remove the $\varepsilon\to0$ divergence in \eqref{diag2comp} we introduce a counterterm by defining, for the bare defect coupling $h_B$,\footnote{We typically do not write the subscript ``$B$'' to indicate bare couplings unless strictly necessary.}
\begin{equation}
    (h_B)_i=\mu^{\varepsilon/2}\Big(h_i+\sum_p(Z_h)_{p,i}\Big)\,,\qquad (Z_h)_{p,i}=\sum_{n=1}^\infty \frac{f_{p,i}^{(n)}}{\varepsilon^n}\,.
\end{equation}
The parameter $p$ indicates the power of the coupling $h$; for example $(Z_h)_{3,i}$ is cubic in $h$. Taking just the $f_{p,i}^{(1)}$ contributions and using \eqref{eqfinite} we find
\begin{equation}\label{count1}
    \begin{tikzpicture}[scale=1.2,baseline=(vert_cent.base)]
        \node (vert_cent) at (0,0.5) {$\phantom{\cdot}$};
        \draw[dashed] (0,1)--(0,0) node[at start,yshift=7pt] {$x,i$};
        \draw[thick] (-0.75,0)--(0.75,0);
        \node[draw,circle,thick,black,cross,fill=white] at (0,0) {};
    \end{tikzpicture}\supset -\frac{1}{4\pi|\mathbf{x}|}\Big(\frac{1}{\varepsilon}+\frac12\big(\gamma+\log(4\pi|\mathbf{x}|^2\mu^2)\big)+\text{O}(\varepsilon)\Big)\sum_{p}f_{p,i}^{(1)}\,.
\end{equation}
Demanding now that
\begin{equation}
    \begin{tikzpicture}[scale=0.8,every node/.style={scale=0.8},baseline=(vert_cent.base)]
        \node (vert_cent) at (0,1) {$\phantom{\cdot}$};
        \draw[dashed] (0,2)--(0,0);
        \draw[thick] (-1.25,0)--(1.25,0);
        \filldraw (0,0) circle (2pt);
        \node [fill, shape=rectangle, minimum width=4pt, minimum height=4pt, inner sep=0pt, anchor=center] at (0,1) {};
        \draw[dashed] (0,1)--(-0.625,0);
        \filldraw (-0.625,0) circle (2pt);
        \draw[dashed] (0,1)--(0.625,0);
        \filldraw (0.625,0) circle (2pt);
    \end{tikzpicture}+
    \begin{tikzpicture}[scale=1.63,baseline=(vert_cent.base)]
        \node (vert_cent) at (0,0.5) {$\phantom{\cdot}$};
        \draw[dashed] (0,1)--(0,0);
        \draw[thick] (-0.5,0)--(0.5,0);
        \node[draw,circle,thick,black,cross,fill=white] at (0,0) {};
    \end{tikzpicture}=\;\text{finite in the limit }\varepsilon\to0\,,
\end{equation}
shows that
\begin{equation}\label{leadZh3}
    f_{3,i}^{(1)}\supset \frac{1}{16\pi^2}\frac{1}{12}\lambda_{ijkl}h_jh_kh_l\,.
\end{equation}

The contribution \eqref{leadZh3} to $f_{3,i}^{(1)}$ leads to a contribution to the beta function of the defect coupling. This is as usual computed by requiring independence of the bare defect coupling $(h_B)_i$ from the renormalisation group scale $\mu$:
\begin{equation}\label{bareindep}
    \mu\frac{d(h_B)_i}{d\mu}=0\quad\Rightarrow\quad \Big(\frac12\lsp\varepsilon+\beta_j\frac{\partial}{\partial h_j}+\beta_{jklm}\frac{\partial}{\partial\lambda_{jklm}}\Big)\Big(h_i+\sum_p(Z_h)_{p,i}\Big)=0\,,
\end{equation}
with $\beta_{ijkl}=-\varepsilon\lambda_{ijkl}+\hat{\beta}_{ijkl}$ the bulk beta function of the quartic coupling, where $\hat{\beta}_{ijkl}$ does not depend explicitly on $\varepsilon$. Eq.~\eqref{bareindep} requires, at order $\varepsilon$,
\begin{equation}\label{defbetagen}
    \beta_i=-\tfrac12\varepsilon\lsp h_i+\hat{\beta}_i\,,
\end{equation}
where $\hat{\beta}_i$ does not depend explicitly on $\varepsilon$. At order $\varepsilon^0$ we can determine $\hat{\beta}_i$ from the residues of the $1/\varepsilon$ poles of $(Z_h)_{p,i}$ only:
\begin{equation}\label{betafromZ}
    \hat{\beta}_i=-\bigg(\frac12\Big(1-h_j\frac{\partial}{\partial h_j}\Big)-\lambda_{jklm}\frac{\partial}{\partial\lambda_{jklm}}\bigg)\sum_p f_{p,i}^{(1)}\,,
\end{equation}
which, from the contribution \eqref{leadZh3}, gives
\begin{equation}\label{defbetahat1}
    \hat{\beta}_i\supset\frac{1}{16\pi^2}\frac16\lsp\lambda_{ijkl}h_jh_kh_l\,.
\end{equation}
Eq.~\eqref{defbetagen} with \eqref{defbetahat1} is the result for the defect beta function at leading order in the bulk coupling.

At next-to-leading order in the bulk coupling we need to consider the last three diagrams of Fig.~\ref{BetaDefDiagScalar2Loops}. Here we have to be mindful of bulk and lower order counterterms: the first two diagrams have counterterms associated with bulk renormalisation, while the last has the counterterm \eqref{count1} with \eqref{leadZh3}.

We have
\begin{equation}\label{scwf1}
    \begin{tikzpicture}[baseline=(vert_cent.base)]
        \node (vert_cent) at (0,1) {$\phantom{\cdot}$};
        \draw[dashed] (0,2)--(0,0);
        \draw[thick] (-1.25,0)--(1.25,0);
        \filldraw (0,0) circle (2pt);
        \node [fill, shape=rectangle, minimum width=4pt, minimum height=4pt, inner sep=0pt, anchor=center] at (0,1.5) {};
        \node [fill, shape=rectangle, minimum width=4pt, minimum height=4pt, inner sep=0pt, anchor=center] at (0,0.5) {};
        \draw[dashed] (0,1) circle (0.5cm);
    \end{tikzpicture}+
    \begin{tikzpicture}[baseline=(vert_cent.base)]
        \node (vert_cent) at (0,1) {$\phantom{\cdot}$};
        \draw[dashed] (0,2)--(0,0);
        \filldraw (0,0) circle (2pt);
        \node[draw,shape=rectangle,minimum width=8.5pt,minimum height=8.5pt,black,cross,fill=white] at (0,1) {};
        \draw[thick] (-1,0)--(1,0);
    \end{tikzpicture}=\frac{1}{1024\pi^5|\mathbf{x}|}\frac{1}{24\lsp\varepsilon}\lambda_{ijkl}\lambda_{jklm}h_m+\text{O}(\varepsilon^0)\,,
\end{equation}
where we used a standard bulk renormalisation result\footnote{With $(\phi_B)_i=(Z_\phi^{1/2})_{ij}(\phi_R)_j$ we have $(Z_\phi)_{ij}=\delta_{ij}-\frac{1}{12\lsp\varepsilon}\lambda_{iklm}\lambda_{jklm}+\ldots$~\cite{Machacek:1983tz, Jack:1983sk}.} and \eqref{eqfinite} to obtain
\begin{equation}
    \begin{tikzpicture}[baseline=(vert_cent.base)]
        \node (vert_cent) at (0,1) {$\phantom{\cdot}$};
        \draw[dashed] (0,2)--(0,0);
        \filldraw (0,0) circle (2pt);
        \node[draw,shape=rectangle,minimum width=8.5pt,minimum height=8.5pt,black,cross,fill=white] at (0,1) {};
        \draw[thick] (-1,0)--(1,0);
    \end{tikzpicture}=-\frac{1}{1024\pi^5|\mathbf{x}|}\frac{1}{24\lsp\varepsilon}\lambda_{ijkl}\lambda_{jklm}h_m+\text{O}(\varepsilon^0)\,.
    \label{scalarbulkrenormgraph}
\end{equation}
The absence of $1/\varepsilon^2$ contributions to \eqref{scwf1} is a reflection of the absence of subdivergences in the two-loop bulk wavefunction renormalisation graph in \eqref{scwf1}. We must require
\begin{equation}
    \begin{tikzpicture}[baseline=(vert_cent.base)]
        \node (vert_cent) at (0,1) {$\phantom{\cdot}$};
        \draw[dashed] (0,2)--(0,0);
        \draw[thick] (-1.25,0)--(1.25,0);
        \filldraw (0,0) circle (2pt);
        \node [fill, shape=rectangle, minimum width=4pt, minimum height=4pt, inner sep=0pt, anchor=center] at (0,1.5) {};
        \node [fill, shape=rectangle, minimum width=4pt, minimum height=4pt, inner sep=0pt, anchor=center] at (0,0.5) {};
        \draw[dashed] (0,1) circle (0.5cm);
    \end{tikzpicture}+
    \begin{tikzpicture}[baseline=(vert_cent.base)]
        \node (vert_cent) at (0,1) {$\phantom{\cdot}$};
        \draw[dashed] (0,2)--(0,0);
        \filldraw (0,0) circle (2pt);
        \node[draw,shape=rectangle,minimum width=8.5pt,minimum height=8.5pt,black,cross,fill=white] at (0,1) {};
        \draw[thick] (-1,0)--(1,0);
    \end{tikzpicture}+
    \begin{tikzpicture}[baseline=(vert_cent.base)]
        \node (vert_cent) at (0,1) {$\phantom{\cdot}$};
        \draw[dashed] (0,2)--(0,0);
        \draw[thick] (-1,0)--(1,0);
        \node[draw,circle,thick,black,cross,fill=white] at (0,0) {};
    \end{tikzpicture}=\;\text{finite in the limit }\varepsilon\to0\,,
\end{equation}
which leads to
\begin{equation}
    f_{1,i}^{(1)}\supset \frac{1}{(16\pi^2)^2}\frac{1}{24}\lambda_{ijkl}\lambda_{jklm}h_m\,.
    \label{scalarrenormfinalresult}
\end{equation}
One sees that, up to a factor of $-4\pi|\mathbf{x}|$ which is due to \eqref{eqfinite}, (\ref{scalarrenormfinalresult}) is predicted by the bulk counterterm (\ref{scalarbulkrenormgraph}). This derives from the fact that the only divergent integrals in diagram (\ref{scwf1}) appear strictly in the bulk, so that the bulk wavefunction renormalisation will already contain all of the information about their divergences. As explored in section \ref{bulkrelation}, this is a general feature of graphs containing only a single defect coupling, which will allow us to easily write down the terms in the beta function linear in $h_i$.

We should also require
\begin{equation}\label{scwf2}
    \begin{tikzpicture}[baseline=(vert_cent.base)]
        \node (vert_cent) at (0,1) {$\phantom{\cdot}$};
        \draw[dashed] (0,2)--(0,1.5);
        \draw[dashed] (0,1.5)--(1,0);
        \draw[thick] (-1.25,0)--(1.25,0);
        \node [fill, shape=rectangle, minimum width=4pt, minimum height=4pt, inner sep=0pt, anchor=center] at (-0.5,0.75) {};
        \draw[dashed] (-0.5,0.75)--(0,0);
        \draw[dashed] (-0.5,0.75)--(-1,0);
        \node [fill, shape=rectangle, minimum width=4pt, minimum height=4pt, inner sep=0pt, anchor=center] at (0,1.5) {};
        \filldraw (0,0) circle (2pt);
        \filldraw (-1,0) circle (2pt);
        \filldraw (1,0) circle (2pt);
        \draw[dashed] (0,1.5) to[out=180,in=120] (-0.5,0.75);
        \draw[dashed] (0,1.5) to[out=285,in=0] (-0.5,0.75);
    \end{tikzpicture}+
    \begin{tikzpicture}[baseline=(vert_cent.base)]
        \node (vert_cent) at (0,1) {$\phantom{\cdot}$};
        \draw[dashed] (0,2)--(0,0);
        \draw[thick] (-1.25,0)--(1.25,0);
        \filldraw (0,0) circle (2pt);
        \draw[dashed] (0,1)--(-0.625,0);
        \filldraw (-0.625,0) circle (2pt);
        \draw[dashed] (0,1)--(0.625,0);
        \filldraw (0.625,0) circle (2pt);
        \node[draw,shape=rectangle,minimum width=8.5pt,minimum height=8.5pt,black,cross,fill=white] at (0,1) {};
    \end{tikzpicture}+
    \begin{tikzpicture}[baseline=(vert_cent.base)]
        \node (vert_cent) at (0,1) {$\phantom{\cdot}$};
        \draw[dashed] (0,2)--(0,0);
        \draw[thick] (-1,0)--(1,0);
        \node[draw,circle,thick,black,cross,fill=white] at (0,0) {};
    \end{tikzpicture}=\;\text{finite in the limit }\varepsilon\to0\,,
\end{equation}
where for the associated bulk subdivergence we have included the well-known bulk counterterm
\begin{equation}
    \begin{tikzpicture}[baseline=(vert_cent.base)]
        \node (vert_cent) at (0,0) {$\phantom{\cdot}$};
        \draw[dashed,shorten <=0.75pt] (-0.5,-0.5)--(0.5,0.5);
        \draw[dashed,shorten <=0.75pt]  (-0.5,0.5)--(0.5,-0.5);
        \node[xshift=-3pt,yshift=3pt] at (-0.5,0.5) {$i$};
        \node[xshift=3pt,yshift=3pt] at (0.5,0.5) {$j$};
        \node[xshift=3pt,yshift=-3pt] at (0.5,-0.5) {$k$};
        \node[xshift=-3pt,yshift=-3pt] at (-0.5,-0.5) {$l$};
        \node[draw,shape=rectangle,minimum width=8.5pt,minimum height=8.5pt,black,cross,fill=white] at (0,0) {};
        \node[xshift=10pt] at (0,0) {};
    \end{tikzpicture}=\frac{1}{16\pi^2}\frac{1}{\varepsilon}(\lambda_{ijmn}\lambda_{mnkl}+\lambda_{ikmn}\lambda_{mnjl}+\lambda_{ilmn}\lambda_{mnjk})\,.
\end{equation}
Eq.\ \eqref{scwf2} requires
\begin{equation}
    f_{3,i}^{(1)}\supset -\frac{1}{(16\pi^2)^2}\frac{1}{12}\lambda_{ijkl}\lambda_{klmn}h_jh_mh_n\,,\qquad f_{3,i}^{(2)}\supset \frac{1}{(16\pi^2)^2}\frac{1}{12}\lambda_{ijkl}\lambda_{klmn}h_jh_mh_n\,.
\end{equation}

Finally, we must demand that
\begin{equation}\label{scwf3}
    \begin{tikzpicture}[baseline=(vert_cent.base)]
        \node (vert_cent) at (0,1) {$\phantom{\cdot}$};
        \draw[dashed] (0,2)--(0,0);
        \draw[thick] (-1.25,0)--(1.25,0);
        \filldraw (0,0) circle (2pt);
        \node [fill, shape=rectangle, minimum width=4pt, minimum height=4pt, inner sep=0pt, anchor=center] at (0,0.7) {};
        \draw[dashed] (0,0.7)--(-0.5,0);
        \filldraw (-0.5,0) circle (2pt);
        \draw[dashed] (0,0.7)--(0.5,0);
        \filldraw (0.5,0) circle (2pt);
        \node [fill, shape=rectangle, minimum width=4pt, minimum height=4pt, inner sep=0pt, anchor=center] at (0,1.4) {};
        \draw[dashed] (0,1.4)--(-1,0);
        \filldraw (-1,0) circle (2pt);
        \draw[dashed] (0,1.4)--(1,0);
        \filldraw (1,0) circle (2pt);
    \end{tikzpicture}+\begin{tikzpicture}[baseline=(vert_cent.base)]
        \node (vert_cent) at (0,1) {$\phantom{\cdot}$};
        \draw[dashed] (0,2)--(0,0);
        \draw[thick] (-1.25,0)--(1.25,0);
        \draw[dashed] (0,1)--(-0.625,0);
        \filldraw (-0.625,0) circle (2pt);
        \draw[dashed] (0,1)--(0.625,0);
        \filldraw (0.625,0) circle (2pt);
        \node [fill, shape=rectangle, minimum width=4pt, minimum height=4pt, inner sep=0pt, anchor=center] at (0,1) {};
        \node[draw,circle,thick,black,cross,fill=white] at (0,0) {};
    \end{tikzpicture}+
    \begin{tikzpicture}[baseline=(vert_cent.base)]
        \node (vert_cent) at (0,1) {$\phantom{\cdot}$};
        \draw[dashed] (0,2)--(0,0);
        \draw[thick] (-1,0)--(1,0);
        \node[draw,circle,thick,black,cross,fill=white] at (0,0) {};
    \end{tikzpicture}=\;\text{finite in the limit }\varepsilon\to0\,,
\end{equation}
where in the middle diagram in the left-hand side we need to use the $1/\varepsilon$ part of the counterterm \eqref{count1} with the lower-order contribution \eqref{leadZh3}. This gives
\begin{equation}
    f_{5,i}^{(1)}\supset -\frac{1}{(16\pi^2)^2}\frac{1}{48}\lambda_{ijkl}\lambda_{jmnp}h_kh_lh_mh_nh_p\,,\qquad
    f_{5,i}^{(2)}\supset \frac{1}{(16\pi^2)^2}\frac{1}{96}\lambda_{ijkl}\lambda_{jmnp}h_kh_lh_mh_nh_p\,.
\end{equation}

The residues of the $1/\varepsilon$ poles determine the beta function according to \eqref{betafromZ}. From \eqref{defbetagen} and including the leading order results \eqref{defbetahat1}, we have, at next-to-leading order,
\begin{equation}\label{BetaDefScalar2Loops}
    \beta_i=-\tfrac12\varepsilon\lsp h_i+\tfrac16\lsp \lambda_{ijkl}h_jh_kh_l+\tfrac{1}{12}\lambda_{ijkl}\lambda_{jklm}h_m-\tfrac14\lambda_{ijkl}\lambda_{klmn}h_jh_mh_n-\tfrac{1}{12}\lambda_{ijkl}\lambda_{jmnp}h_kh_lh_mh_nh_p\,,
\end{equation}
where we have rescaled $\lambda\to 16\pi^2\lsp\lambda$. Eq.~\eqref{BetaDefScalar2Loops} agrees with the results of \cite{Cuomo:2021kfm} for the case of the $O(N)$ model.\footnote{To obtain the $O(N)$ model we need to set $\lambda_{ijkl}=\lambda(\delta_{ij}\delta_{kl}+\delta_{ik}\delta_{jl}+\delta_{il}\delta_{jk})$.} It is also consistent with the results of \cite{Allais:2014fqa}, up to a scheme change in \eqref{scwf2}. Finally, it is consistent with the general results of \cite{Rodriguez-Gomez:2022gbz} and \cite{Popov:2022nfq}.

\subsection{Scalars and fermions in the bulk}
When fermions are included in the bulk, the relevant action is
\begin{equation}\label{scferLag}
S=\int d^{\lsp d}x\,\big(\tfrac12\lsp\partial^\mu\phi_i\lsp\partial_\mu\phi_i+i\bar{\psi}_a\bar{\sigma}^\mu\partial_\mu\psi_a+\tfrac{1}{4!}\lambda_{ijkl}\lsp\phi_i\phi_j\phi_k\phi_l+(\tfrac12\lsp  y_{iab}\lsp\phi_i\psi_a\psi_b+\text{h.c.})\big)\,,
\end{equation}
where $\phi_i, i=1,\ldots,N_s$ are again real scalar fields and $\psi_a, a=1,\ldots,N_f$, are two-component fermions. The coupling tensor $y_{iab}$ is symmetric in the fermionic flavour indices. In the presence of fermions renormalisation of the scalar propagator requires $(\phi_B)_i=(Z_\phi^{1/2})_{ij}(\phi_R)_j$ with~\cite{Machacek:1983tz, Jack:1984vj}
\begin{equation}\label{Zphi}
    (Z_\phi)_{ij}=\delta_{ij}-\frac{1}{\varepsilon}\big(Y_{ij}-\tfrac12\lsp\widetilde{Y}_{ikjk}-\tfrac34\lsp\widetilde{Y}_{ikkj}+\tfrac{1}{12}\lambda_{iklm}\lambda_{jklm}\big)+\ldots\,,
\end{equation}
where
\begin{equation}
    \begin{aligned}
        Y_{ij}&=y_{iab}y^*{\!\!\!}_{jab}+y^*{\!\!\!}_{iab}y_{jab}= \Tr(y_iy^*{\!\!\!}_j+y^*{\!\!\!}_iy_j)\,,\\
        \widetilde{Y}_{ijkl}&=\Tr(y_iy^*{\!\!\!}_jy_ky^*{\!\!\!}_l+y^*{\!\!\!}_iy_jy^*{\!\!\!}_ky_l)\,.
    \end{aligned}
\end{equation}
We have included the $1/\varepsilon$ results up to two loops and rescaled $y\to 4\pi\lsp y$, $\lambda\to 16\pi^2\lsp\lambda$.

The contributions to $\langle\phi_i(x)\rangle$ that involve fermions are described by the diagrams in Fig.~\ref{BetaDefScalarFermionDiag}. The computation proceeds in a straightforward way, which we describe in Appendix \ref{app:fermions}. The most difficult graph is the fourth one, whose $1/\varepsilon$ pole can however be extracted from \cite{Giombi:2022vnz}.
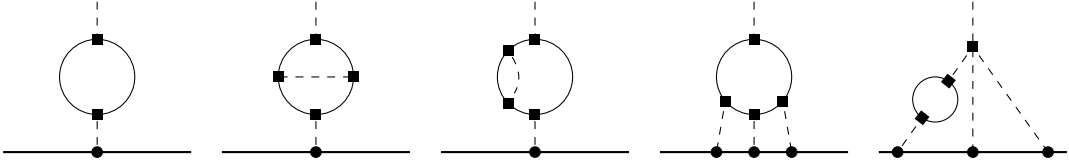
\begin{figure}[ht]
\centering
    \begin{tikzpicture}[baseline=(vert_cent.base)]
        \node (vert_cent) at (0,1) {$\phantom{\cdot}$};
        \draw[dashed] (0,2)--(0,1.5);
        \draw[dashed,shorten <=2.5pt] (0,0.5)--(0,0);
        \draw[thick] (-1.25,0)--(1.25,0);
        \filldraw (0,0) circle (2pt);
        \node [fill, shape=rectangle, minimum width=4pt, minimum height=4pt, inner sep=0pt, anchor=center] at (0,1.5) {};
        \node [fill, shape=rectangle, minimum width=4pt, minimum height=4pt, inner sep=0pt, anchor=center] at (0,0.5) {};
        \draw (0,1) circle (0.5cm);
    \end{tikzpicture}\quad
    \begin{tikzpicture}[baseline=(vert_cent.base)]
        \node (vert_cent) at (0,1) {$\phantom{\cdot}$};
        \draw[dashed] (0,2)--(0,1.5);
        \draw[dashed,shorten <=2.5pt] (0,0.5)--(0,0);
        \draw[thick] (-1.25,0)--(1.25,0);
        \filldraw (0,0) circle (2pt);
        \node [fill, shape=rectangle, minimum width=4pt, minimum height=4pt, inner sep=0pt, anchor=center] at (0,1.5) {};
        \node [fill, shape=rectangle, minimum width=4pt, minimum height=4pt, inner sep=0pt, anchor=center] at (0,0.5) {};
        \node [fill, shape=rectangle, minimum width=4pt, minimum height=4pt, inner sep=0pt, anchor=center] at (-0.5,1) {};
        \node [fill, shape=rectangle, minimum width=4pt, minimum height=4pt, inner sep=0pt, anchor=center] at (0.5,1) {};
        \draw[dashed,shorten <=0.5pt] (-0.5,1)--(0.5,1);
        \draw (0,1) circle (0.5cm);
    \end{tikzpicture}\quad
    \begin{tikzpicture}[baseline=(vert_cent.base)]
        \node (vert_cent) at (0,1) {$\phantom{\cdot}$};
        \draw[dashed] (0,2)--(0,1.5);
        \draw[dashed,shorten <=2.5pt] (0,0.5)--(0,0);
        \draw[thick] (-1.25,0)--(1.25,0);
        \filldraw (0,0) circle (2pt);
        \node [fill, shape=rectangle, minimum width=4pt, minimum height=4pt, inner sep=0pt, anchor=center] at (0,1.5) {};
        \node [fill, shape=rectangle, minimum width=4pt, minimum height=4pt, inner sep=0pt, anchor=center] at (0,0.5) {};
        \draw[white] (0,1)--++(135:0.5) node (u) {};
        \draw[white] (0,1)--++(225:0.5) node (d) {};
        \node [fill, shape=rectangle, minimum width=4pt, minimum height=4pt, inner sep=0pt, anchor=center] at (u) {};
        \node [fill, shape=rectangle, minimum width=4pt, minimum height=4pt, inner sep=0pt, anchor=center] at (d) {};
        \draw[dashed,shorten <=-1.5pt] (u) to[out=-60,in=60] (d);
        \draw (0,1) circle (0.5cm);
    \end{tikzpicture}\quad
    \begin{tikzpicture}[baseline=(vert_cent.base)]
        \node (vert_cent) at (0,1) {$\phantom{\cdot}$};
        \draw[dashed] (0,2)--(0,1.5);
        \draw[dashed,shorten <=2.5pt] (0,0.5)--(0,0);
        \draw[thick] (-1.25,0)--(1.25,0);
        \filldraw (0,0) circle (2pt);
        \filldraw (-0.5,0) circle (2pt);
        \filldraw (0.5,0) circle (2pt);
        \node [fill, shape=rectangle, minimum width=4pt, minimum height=4pt, inner sep=0pt, anchor=center] at (0,1.5) {};
        \node [fill, shape=rectangle, minimum width=4pt, minimum height=4pt, inner sep=0pt, anchor=center] at (0,0.5) {};
        \draw[white] (0,1)--++(320:0.5) node (r) {};
        \draw[white] (0,1)--++(220:0.5) node (l) {};
        \node [fill, shape=rectangle, minimum width=4pt, minimum height=4pt, inner sep=0pt, anchor=center] at (l) {};
        \node [fill, shape=rectangle, minimum width=4pt, minimum height=4pt, inner sep=0pt, anchor=center] at (r) {};
        \draw (0,1) circle (0.5cm);
        \draw[dashed] (l)--(-0.5,0);
        \draw[dashed] (r)--(0.5,0);
    \end{tikzpicture}\quad
    \begin{tikzpicture}[baseline=(vert_cent.base)]
        \node (vert_cent) at (0,1) {$\phantom{\cdot}$};
        \draw[dashed] (0,2)--(0,0);
        \draw[thick] (-1.25,0)--(1.25,0);
        \filldraw (0,0) circle (2pt);
        \draw[white,name path=LP] (0,1.4)--(-1,0) node[midway] (l) {};
        \node [fill, shape=rectangle, minimum width=4pt, minimum height=4pt, inner sep=0pt, anchor=center] at (0,1.4) {};
        \draw[dashed,shorten <=-2pt] (0,1.4)--(1,0);
        \filldraw (1,0) circle (2pt);
        \filldraw (-1,0) circle (2pt);
        \draw[fill=white,name path=CIR] (l) circle (0.3cm);
        \node [name intersections={of=LP and CIR}, fill, shape=rectangle, rotate=52, minimum width=4pt, minimum height=4pt, inner sep=0pt, anchor=center] at (intersection-1) {};
        \node [name intersections={of=LP and CIR}, fill, shape=rectangle, rotate=52, minimum width=4pt, minimum height=4pt, inner sep=0pt, anchor=center] at (intersection-2) {};
        \draw[dashed,name intersections={of=LP and CIR},shorten <=-2pt] (0,1.4)--(intersection-1);
        \draw[dashed,name intersections={of=LP and CIR},shorten <=-2pt] (intersection-2)--(-1,0);
    \end{tikzpicture}
    \caption{Diagrams that contribute to the beta function of the defect coupling involving fermions in the bulk, up to next-to-leading order.}
    \label{BetaDefScalarFermionDiag}
\end{figure}
Along with the purely-scalar contributions \eqref{BetaDefScalar2Loops}, the defect coupling beta function at next-to-leading order is
\begin{equation}\label{BetaDefScalarFermion}
\begin{aligned}
        \beta_i&=-\tfrac12\varepsilon\lsp h_i+\tfrac16\lsp \lambda_{ijkl}h_jh_kh_l+\tfrac{1}{12}\lambda_{ijkl}\lambda_{jklm}h_m-\tfrac14\lambda_{ijkl}\lambda_{klmn}h_jh_mh_n-\tfrac{1}{12}\lambda_{ijkl}\lambda_{jmnp}h_kh_lh_mh_nh_p\\
        &\quad\,+\tfrac12\lsp Y_{ij}h_j-\tfrac14\lsp \widetilde{Y}_{ijkj}h_k - \tfrac38\lsp \widetilde{Y}_{ijjk}h_k+(1-\tfrac{1}{6}\pi^2)\widetilde{Y}_{ijkl}h_jh_kh_l -\tfrac{1}{4}\lsp \lambda_{ijkl}Y_{lm}h_jh_kh_m\,.
    \end{aligned}
\end{equation}

\subsection{Relation to bulk wavefunction renormalisation}\label{bulkrelation}
Coefficients of terms in $\beta_i$ linear in $h$ are simply given by the coefficients of the corresponding diagrams that renormalise the scalar propagator in the bulk. Indeed, renormalisation of the scalar two-point function in the bulk requires $(Z_\phi)_{ij}=\delta_{ij}+\sum_{n=1}^{\infty}b_{ij}^{(n)}/\varepsilon^n$, and then $(\gamma_\phi)_{ij}=-\frac12\rho_{I}g_I\lsp\partial b^{(1)}_{ij}/\partial g_I$, where $g_I$ stands for either $\lambda_{ijkl}$, $y_{iab}$ or $y^*_{iab}$, and $\rho_I$ is equal to 1 when its index corresponds to a quartic coupling and $\frac12$ when it corresponds to a Yukawa coupling. For example, $(\gamma_\phi)_{ij}\supset\frac12 Y_{ij}$. The same scalar wavefunction renormalisation appears in the defect computation, with which we renormalise $h_i$.

From Eq.\ (\ref{eqfinite}) we see that there is no divergence associated with attaching a propagator to the defect. Therefore, any divergence in these sort of diagrams must arise purely from the divergence associated with renormalisation in the bulk. These divergences will thus be exactly cancelled by the bulk counterterms. However, as we are only considering the one-point function we should not subtract the full propagator counterterm but only half of it, leaving the other half for the defect counterterm. That is to say, for renormalisation linear in $h$ we get
\begin{equation}
    (Z_h)_{1,i}=\begin{tikzpicture}[baseline=(vert_cent.base)]
        \node (vert_cent) at (0,1) {$\phantom{\cdot}$};
        \draw[dashed] (0,2)--(0,0);
        \draw[thick] (-1.25,-0.1)--(1.25,-0.1);
        \node[draw, fill=gray!20, shape=rectangle, minimum width=0.75cm, minimum height=0.75cm, inner sep=0pt, anchor=center] at (0,1) {};
        \filldraw (0,-0.1) circle (2pt) node[below] {$h$};
    \end{tikzpicture}=\tfrac12(h_i-(Z_\phi)_{ij}h_{j})\,.
    \label{fieldcorrections}
\end{equation}
Thus, the computation of $(Z_\phi)_{ij}$ in the bulk, e.g.\ \eqref{Zphi} for scalar-fermion theories up to two loops, allows us to easily determine a subset of the contributions to the beta functions of the defect couplings. Using the above definition of the anomalous dimension for the scalar field, one finds that the diagrams (\ref{fieldcorrections}) contribute a term $\frac{1}{2}(\gamma_\phi)_{ij}h_j$. One can then combine this term with the classical beta function to find that the beta function takes the form
\begin{equation}
    \beta_i=\big(\!-\!\tfrac12\varepsilon\lsp\delta_{ij}+(\gamma_\phi)_{ij}\big)h_j+\cdots=\big((\Delta_\phi)_{ij}-\delta_{ij}\big)h_j+\cdots\,,
\end{equation}
where the ellipses represent terms of higher order in the defect coupling arising from diagrams with more scalar legs attaching to the defect.

\section{Scalar fields}\label{scalarsec}
For scalar vector models described by \eqref{laggen}, since $\Delta_\phi=1-\frac12\varepsilon$ at leading order in $\varepsilon$ the defect deformation is relevant. We will consider only the leading order results for the defect beta function obtained above. At leading order the flow is gradient, as there exists a quantity $H$ given by
\begin{equation}
    H=-\tfrac14\varepsilon\lsp h^2+\tfrac{1}{24}\lsp\lambda_{ijkl}h_ih_jh_kh_l\,,\qquad h^2=h_ih_i\,,
\end{equation}
such that
\begin{equation}
    \beta_i=\frac{\partial H}{\partial h_i}\,.
\end{equation}
We have $H=2\log g$, where $\log g$ is discussed in \cite{Cuomo:2021rkm, Cuomo:2021kfm}.

To consider dCFTs, we take the bulk theory \eqref{laggen} to be critical, meaning that
\begin{equation}\label{fpbulk}
    \varepsilon\lsp\lambda_{ijkl}=\lambda_{ijmn}\lambda_{klmn}+\lambda_{ikmn}\lambda_{jlmn}+\lambda_{ilmn}\lambda_{jkmn}\,,
\end{equation}
and look for non-trivial fixed points of
\begin{equation}\label{betahgen}
    \beta_i=-\tfrac12\varepsilon\lsp h_i+\tfrac16\lambda_{ijkl}h_jh_kh_l
\end{equation}
in the space of the $h_i$ couplings.

If one starts with the free bulk theory, for which $\lambda_{ijkl}=0$, then it is obvious that the only root of \eqref{betahgen} is $h_i=0$. In that case, $H=0$. Now consider a non-trivial bulk CFT. The $h_i=0$ root of \eqref{betahgen} remains, and again $H=0$, but now we may also seek non-trivial roots for which $H\neq0$ if $h_i$ are real. At a non-trivial dCFT, then, we may use
\begin{equation}\label{betaihi}
    \beta_i h_i=0\quad\Rightarrow\quad \lambda_{ijkl}h_ih_jh_kh_l=3\lsp\varepsilon\lsp h^2
\end{equation}
to obtain
\begin{equation}
    H=-\tfrac18\varepsilon\lsp h^2\leq0\,.
\end{equation}
Thus, any non-trivial dCFT that may arise in the IR by deforming the bulk theory will necessarily have $H<0$ and the $g$-theorem~\cite{Cuomo:2021rkm} will be satisfied. As we will see below there may exist bulk theories in which multiple inequivalent dCFTs can be found. In those cases the $g$-theorem predicts that the stable dCFT is the one with the smallest value of $H$. Our results are consistent with this. We also find examples of multiple stable dCFTs, in which $H$ is different. Nevertheless, there are no RG flows connecting these stable dCFTs.

We may consider as special cases bulk vector models obtained by single-coupling deformations of the $O(N)$ theory:
\begin{equation}\label{lag}
    S=\int d^{\lsp d}x\,\big(\tfrac12\partial^\mu\phi_i \lsp\partial_\mu\phi_i+\tfrac18\lambda\lsp(\phi^2)^2+\tfrac{1}{24}\lsp g\lsp d_{ijkl}\phi_i\phi_j\phi_k\phi_l\big)\,.
\end{equation}
Such deformations break $O(N)$ to a subgroup $G$, and $d_{ijkl}$ is a rank-four invariant tensor of $G$ that is symmetric and traceless. It is well-known that beyond the free and $O(N)$ theories the action \eqref{lag} has two further fixed points with global symmetry $G$ given by
\begin{equation}\label{fpcoup}
    \lambda_{\pm}=\frac{1}{N+8+X_{\pm}{\!}^2}\lsp\varepsilon\,,\qquad 
    \sqrt{a}\,g_\pm=\frac{X_{\pm}}{N+8+X_{\pm}{\!}^2}\lsp\varepsilon\,,
\end{equation}
where $X_{\pm}=\frac12\big(3\lsp b/\sqrt{a}\pm\sqrt{16-4N+9\lsp b^2/a}\big)$. The parameters $a$ and $b$ are determined by~\cite{Osborn:2017ucf}
\begin{equation}\label{ddeq}
    d_{ijmn}\lsp d_{klmn}=\tfrac{1}{N-1}a\big(\tfrac12 N(\delta_{ik}\delta_{jl}+\delta_{il}\delta_{jk})-\delta_{ij}\delta_{kl}\big)+e^u\lsp w_{u,ijkl}+b\, d_{ijkl}\,,
\end{equation}
where $w_{u,ijkl}$ are potential further rank-four invariant tensors of the group $G$ (the index $u$ simply counts such invariant tensors) satisfying
\begin{equation}
w_{u,ijkl} = w_{u,ji\hskip 0.5ptkl}  = w_{u,kl\hskip 0.5 pt ij} \, , \qquad w_{u,i(jkl)} = 0 \, ,
\qquad  w_{u,ii\hskip 0.5 pt kl} =0 \, .
\end{equation}
The relation \eqref{ddeq} ensures that the RG flow is restricted to the space of the two couplings $\lambda$ and $g$.

For the general line defect deformation of \eqref{lag} the beta function of $h_i$ follows from \eqref{betahgen} and reads
\begin{equation}\label{betah}
    \beta_i=-\tfrac12\varepsilon\lsp h_i+\tfrac12(\lambda\lsp h_i\lsp h^2+\tfrac13\lsp g\, d_{ijkl}h_jh_kh_l)\,.
\end{equation}
This is the result at leading order in $\lambda, g$. 

As an aside let us note here that in multiscalar models of the type \eqref{lag} the $\phi^2$ operator has anomalous dimension equal to $(N+2)\lambda$, where $\lambda$ is the fixed point value of the coupling given by \eqref{fpcoup}. Its scaling dimension is thus $\Delta_{\phi^2}=2-\varepsilon+(N+2)\lambda$ and we find that $\Delta_{\phi^2}<2$ for $\lambda<\frac{1}{N+2}\varepsilon$. The solutions in \eqref{fpcoup} indeed satisfy $\lambda_\pm<\frac{1}{N+2}\varepsilon$, and one may thus consider surface defect deformations of these theories, with $\phi^2$ localised on a surface as the symmetry-preserving perturbing operator. Operators of the type $\phi_i\phi_j$ need to be decomposed under the global symmetry preserved by the bulk CFT. The associated operators in the appropriate irreducible representations under the symmetry of the bulk CFT can be used as symmetry-breaking surface defect deformations if their dimension is below 2, but results here need to be discussed in a case by case basis.\footnote{Surface defects in $d=6-\varepsilon$ have recently been discussed in \cite{Rodriguez-Gomez:2022gbz}.}

In the remainder of this section we will analyse in detail defect deformations in a few examples. The bulk CFTs we will be perturbing around have been discussed in detail in \cite{Osborn:2017ucf, Rychkov:2018vya}.

\subsection{\texorpdfstring{$O(N)$}{O(N)} model}
In the simple case of the $O(N)$ model we have
\begin{equation}
    \lambda=\frac{1}{N+8}\varepsilon\,,\qquad g=0\,,
\end{equation}
for the bulk theory \eqref{lag}, and then
\begin{equation}
    \beta_{i}=-\tfrac12\varepsilon\lsp h_i\big(1-\tfrac{1}{N+8}h^2\big)\,.
\end{equation}
If $h_i\neq 0$ we find that $\beta_i=0$ for
\begin{equation}\label{solON}
    h^2=N+8\,.
\end{equation}
Note that the individual $h_i$'s are left undetermined but are subject to the constraint \eqref{solON}, i.e.\ they live on an $(N-1)$-dimensional sphere.

Looking at the stability matrix $\partial_i\beta_j$ evaluated at \eqref{solON} we find that the operator $\mathcal{O}=h_i\phi_i$ has dimension $1+\varepsilon$, while $N-1$ operators that can be chosen to have the form $\mathcal{O}_{\hat{\imath}}=h_1\phi_{\hat{\imath}}-h_{\hat{\imath}}\phi_1\,,{\hat{\imath}}=2,\ldots,N$ have dimension exactly 1. These operators transform in the vector representation of $O(N-1)$, and we see that the dCFT breaks $O(N)$ to $O(N-1)$. The quotient $O(N)/O(N-1)$ is isomorphic to $S^{N-1}$. Operators like $\mathcal{O}_{\hat{\imath}}$ are sometimes called tilt operators in the literature. 

We would like to emphasise here that the $\mathcal{O}_i$'s do not generate non-trivial deformations of the dCFT defined by \eqref{solON}. All dCFTs on the hypersphere defined by \eqref{solON} are physically equivalent, in the sense that local CFT data do not depend on the specific $h_i$ that satisfies \eqref{solON}. Despite the fact that the quotient is trivial in this sense, its presence implies that a certain combination of integrated connected four-point functions involving the $\mathcal{O}_i$'s corresponds to its Riemann curvature~\cite{Drukker:2022pxk}.

\subsection{Hypercubic model}\label{hypercubic}
For the hypercubic model with global symmetry $B_N=\mathbb{Z}_2{\!}^N\rtimes S_N$, where $S_N$ is the group of permutations of $N$ objects, we have
\begin{equation}
    d_{ijkl}\phi_i\phi_j\phi_k\phi_l=\sum_i\phi_i^{\,4}-\frac{3}{N+2}(\phi^2)^2
    \label{eq:cubd}
\end{equation}
in \eqref{lag}, and a non-trivial fixed point is found for
\begin{equation}
    \lambda=\frac{2(N-1)}{3N(N+2)}\varepsilon\,,\qquad g=\frac{N-4}{3N}\varepsilon\,.
\end{equation}
The other fixed point in \eqref{fpcoup} corresponds to $N$ decoupled Ising models. The defect beta function is
\begin{equation}
    \beta_i=-\tfrac12\varepsilon\lsp h_i\big(1-\tfrac{1}{3N}h^2-\tfrac{N-4}{9N}h_i^2\big)\,.
\end{equation}
The equation $\beta_i=0$ has $1+\sum_{n=1}^N \binom{N}{n} 2^n=3^N$ solutions, falling into $N+1$ equivalence classes preserving different global symmetries on the defect. One of them is the solution $h_i=0$. The other $N$ equivalence classes of solutions are given by choosing $n$ couplings to be equal to each other in absolute value so that their squares are all equal to $\hat{h}_n^2$ and $N-n$ couplings to be zero, for $n=1,\ldots,N$. Then, the beta functions corresponding to couplings that were set to zero are also zero trivially, while the remaining ones satisfy $\beta_{\hat{\imath}}/h_{\hat{\imath}}=-\frac12\varepsilon(1-\frac{n}{3N}\hat{h}_n^2-\frac{N-4}{9N}\hat{h}_n^2)$, which becomes zero for 
\begin{equation}
    \hat{h}_n^2=\frac{9N}{N+3n-4}\,.
\end{equation}
The dCFT with $n$ couplings non-zero has global symmetry $B_{N-n}\times S_n$.

The stability matrix $S_{ij}=\partial_i\beta_j$ takes a block-diagonal form,
\begin{equation}
    S=\begin{pmatrix}
        P & 0\\
        0 & Q
    \end{pmatrix}\,.
\end{equation}
Corresponding to the $n$ non-zero couplings we have the $n\times n$ matrix
\begin{equation}
P=\tfrac{N-4}{N+3n-4}\varepsilon \begin{pmatrix}
1 & 0 & \cdots & 0\\
0 & 1 & \cdots & 0\\
\vdots & \vdots  & \ddots & \vdots\\
0 & 0 & \cdots & 1\\
\end{pmatrix}
+\tfrac{3}{N+3n-4}\varepsilon
\begin{pmatrix}
1 & 1 & \cdots & 1\\
1 & 1 & \cdots & 1\\
\vdots & \vdots  & \ddots & \vdots\\
1 & 1 & \cdots & 1\\
\end{pmatrix}\,,
\end{equation}
while for the $N-n$ zero couplings we have the $(N-n)\times(N-n)$ multiple of the identity matrix
\begin{equation}
Q=-\tfrac{N-4}{2(N+3n-4)}\varepsilon \begin{pmatrix}
1 & 0 & \cdots & 0\\
0 & 1 & \cdots & 0\\
\vdots & \vdots  & \ddots & \vdots\\
0 & 0 & \cdots & 1
\end{pmatrix}\,.
\end{equation}

When $n=N$ the eigenvalues of the stability matrix give one operator with dimension $1+\varepsilon$ and $N-1$ operators with dimension $1+\frac{N-4}{4(N-1)}\varepsilon$. Obviously the case $n=N$ corresponds to the unique stable fixed point when $N\geq4$. The corresponding IR dCFT has global symmetry $S_N$. All other fixed points we have found for $N\geq4$ and up to $N=9$ are unstable. The case $N=4$ is special as then we have coincidence with the $O(4)$ case.\foot{This coincidence does not persist beyond the leading loop order.} 

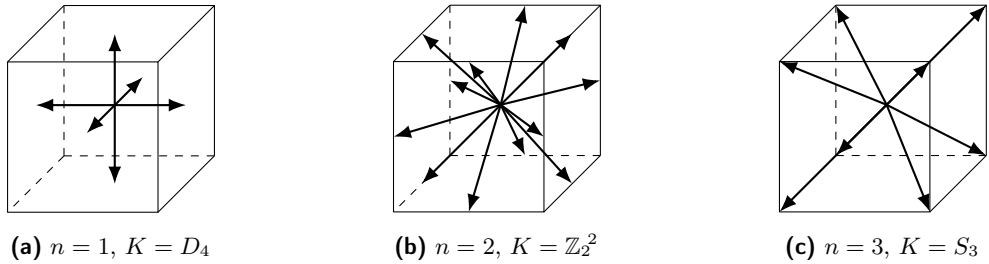
\begin{figure}[ht]
    \centering
    \begin{subfigure}[b]{0.3\textwidth}
    \centering
    \begin{tikzpicture}
        \draw (0,0)--(0,2)--(2,2)--(2,0)--(0,0);
        \draw[xshift=0.75cm,yshift=0.75cm] (2,0)--(2,2)--(0,2);
        \draw (2.75,2.75)--(2,2);
        \draw (0.75,2.75)--(0,2);
        \draw (2,0)--(2.75,0.75);
        \draw[dashed] (0.75,0.75)--(0.75,2.75);
        \draw[dashed] (0.75,0.75)--(0,0);
        \draw[dashed] (0.75,0.75)--(2.75,0.75);
        \draw[-Latex,thick] (1.425,1.425)--(1.425,2.375);
        \draw[-Latex,thick] (1.425,1.425)--(1.425,0.375);
        \draw[-Latex,thick] (1.425,1.425)--(0.375,1.425);
        \draw[-Latex,thick] (1.425,1.425)--(2.375,1.425);
        \draw[-Latex,thick] (1.425,1.425)--(1.8,1.8);
        \draw[-Latex,thick] (1.425,1.425)--(1.05,1.05);
    \end{tikzpicture}
    \caption{$n=1$, $K=D_4$}
    \end{subfigure}
    \begin{subfigure}[b]{0.3\textwidth}
    \centering
    \begin{tikzpicture}
        \draw (0,0)--(0,2)--(2,2)--(2,0)--(0,0);
        \draw[xshift=0.75cm,yshift=0.75cm] (2,0)--(2,2)--(0,2);
        \draw (2.75,2.75)--(2,2);
        \draw (0.75,2.75)--(0,2);
        \draw (2,0)--(2.75,0.75);
        \draw[dashed] (0.75,0.75)--(0.75,2.75);
        \draw[dashed] (0.75,0.75)--(0,0);
        \draw[dashed] (0.75,0.75)--(2.75,0.75);
        \draw[-Latex,thick] (1.425,1.425)--(2,1);
        \draw[-Latex,thick] (1.425,1.425)--(0.75,1.75);
        \draw[-Latex,thick] (1.425,1.425)--(2.75,1.75);
        \draw[-Latex,thick] (1.425,1.425)--(0,1);
        \draw[-Latex,thick] (1.425,1.425)--(0.375,0.375);
        \draw[-Latex,thick] (1.425,1.425)--(1,0);
        \draw[-Latex,thick] (1.425,1.425)--(2.375,0.375);
        \draw[-Latex,thick] (1.425,1.425)--(1.75,0.75);
        \draw[-Latex,thick] (1.425,1.425)--(1,2);
        \draw[-Latex,thick] (1.425,1.425)--(2.375,2.375);
        \draw[-Latex,thick] (1.425,1.425)--(1.75,2.75);
        \draw[-Latex,thick] (1.425,1.425)--(0.375,2.375);
    \end{tikzpicture}
    \caption{$n=2$, $K=\mathbb{Z}_2{\!}^2$}
    \end{subfigure}
    \begin{subfigure}[b]{0.3\textwidth}
    \centering
    \begin{tikzpicture}
        \draw (0,0)--(0,2)--(2,2)--(2,0)--(0,0);
        \draw[xshift=0.75cm,yshift=0.75cm] (2,0)--(2,2)--(0,2);
        \draw (2.75,2.75)--(2,2);
        \draw (0.75,2.75)--(0,2);
        \draw (2,0)--(2.75,0.75);
        \draw[dashed] (0.75,0.75)--(0.75,2.75);
        \draw[dashed] (0.75,0.75)--(0,0);
        \draw[dashed] (0.75,0.75)--(2.75,0.75);
        \draw[-Latex,thick] (1.425,1.425)--(0,0);
        \draw[-Latex,thick] (1.425,1.425)--(0,2);
        \draw[-Latex,thick] (1.425,1.425)--(2,2);
        \draw[-Latex,thick] (1.425,1.425)--(2,0);
        \draw[-Latex,thick] (1.425,1.425)--(0.75,0.75);
        \draw[-Latex,thick] (1.425,1.425)--(2.75,2.75);
        \draw[-Latex,thick] (1.425,1.425)--(0.75,2.75);
        \draw[-Latex,thick] (1.425,1.425)--(2.75,0.75);
    \end{tikzpicture}
    \caption{$n=3$, $K=S_3$}
    \end{subfigure}
    \caption{The three symmetry breaking patterns for $N=3$. $K$ is the subgroup of $B_3$ preserved in each case. The vectors $h_i$ drawn point to the center of faces (left), center of edges (middle) and vertices (right) of the cube.}
    \label{cubicbr}
\end{figure}

For $n=1$ $S$ is obviously diagonal and has one eigenvalue equal to $\varepsilon$ and $N-1$ eigenvalues equal to $-\frac{N-4}{2(N-1)}$. For $N=3$ these latter eigenvalues are positive and thus in that case a stable fixed point is the $n=1$ one and it turns out to be the only fixed point with that property. The global symmetry of the IR stable dCFT for this $N=3$ case is the dihedral group $D_4$ of order 8. In Fig.~\ref{cubicbr} we draw the vectors $h_i$ in the quotient $B_3/K$ for the three different non-trivial dCFTs obtained in this case.

\subsection{Hypertetrahedral model}
To describe a theory with hypertetrahedral symmetry $T_N=S_{N+1}\times\mathbb{Z}_2$, we introduce $N+1$ vectors in $N$-space, $(e_N)_i^\alpha$, $i=1,\ldots,N$, $\alpha=1,\ldots,N+1$, which give the locations of the $N+1$ vertices of an $N$-dimensional hypertetrahedron. Starting from $N=1$ with $(e_1)_1^1=-(e_1)_1^2=-\frac{1}{\sqrt{2}}$, we define, recursively,
\begin{equation}
\begin{aligned}
    (e_N)_i^\alpha &= (e_{N-1})_i^\alpha\,, \qquad i=1,\ldots,N-1,\;\alpha=1,\ldots,N\,,\\
    (e_N)_N^\alpha&=-\sqrt{\frac{1}{N(N+1)}}\,,\qquad \alpha=1,\ldots, N\,,\\
    (e_N)_i^{N+1}&=\sqrt{\frac{N}{N+1}}\delta_i{\hspace{-0.4pt}}^N\,.
\end{aligned}
\end{equation}
These vectors satisfy
\begin{equation}
    \sum_{\alpha} (e_N)_i^\alpha=0\,,\qquad
    \sum_{\alpha}(e_N)_i^\alpha(e_N)_j^\alpha=\delta_{ij}\,,\qquad
    (e_N)_i^\alpha(e_N)_i^\beta=\delta^{\alpha\beta}-\frac{1}{N+1}\,,
    \label{eq:evectorrules}
\end{equation}
and they define a hypertetrahedron of edge length $\sqrt{2}$, with its associated ciscumscribed hypersphere having radius $\sqrt{\frac{N}{N+1}}$. The bulk theory is then given by \eqref{lag} with
\begin{equation}
    d_{ijkl}\phi_i\phi_j\phi_k\phi_l=\sum_\alpha \big((e_N)_i^\alpha\phi_i\big)^4-\frac{3N}{(N+1)(N+2)}(\phi^2)^2\,.
    \label{eq:tetd}
\end{equation}

There are two fixed points with hypertetrahedral symmetry, with couplings given by
\begin{equation}
    \lambda_-=\frac{2(N+1)}{3(N+2)(N+3)}\varepsilon\,,\qquad g_-=\frac{N+1}{3(N+3)}\varepsilon\,,
    \label{eq:tetfixedpointnegative}
\end{equation}
and
\begin{equation}
    \lambda_+=\frac{(N-1)(N-2)}{3(N+2)(N^2-5N+8)}\varepsilon\,,\qquad g_+=\frac{(N-4)(N+1)}{3(N^2-5N+8)}\varepsilon\,.
    \label{eq:tetfixedpointpositive}
\end{equation}
We will refer to the former as $T_{N\lsp -}$ and the latter as $T_{N\lsp +}$. For $N\leq4$ one or both of these coincide with other fixed points. The first non-trivial hypertetrahedral fixed point is $T_{4\lsp -}$, while for $N=5$ the $T_{5\lsp\pm}$ fixed points coincide. For $N>5$ there are distinct hypertetrahedral fixed points at leading order in the $\varepsilon$ expansion.

To discuss a general line defect deformation we use
\begin{equation}
    d_{ijkl}h_jh_kh_l=\sum_{\alpha}(e_N)_i^\alpha \big((e_N)_j^\alpha h_j\big)^3-\frac{3N}{(N+1)(N+2)}h^2h_i
\end{equation}
in \eqref{betah}. Explicit forms of the beta functions are rather unsightly, and an analysis for general $N$ appears to be complicated, due to the presence of the $e_N$ vectors. However, one can overcome this difficulty by noticing that there is a correspondence between $T_N$ and the hypercubic system with one more field, $C_{N+1}$. If we consider the $N+1$ fields
\begin{equation}
    \phi^\alpha=(e_N)^{\alpha}_i\phi_i\,,
\end{equation}
then using the properties \eqref{eq:evectorrules} of the $e_N$ vectors one can see that (\ref{eq:tetd}) becomes
\begin{equation}
    d_{ijkl}\phi_i\phi_j\phi_k\phi_l=\sum_{\alpha}(\phi^{\alpha})^4-\frac{3N}{(N+1)(N+2)}(\phi^2)^2\,,\qquad \phi^2=\phi^\alpha\phi^\alpha\,,
\end{equation}
which, up to the coefficient of the second term in the right-hand side, is of the form of the rank-four tensor of the hypercubic case with $N+1$ fields given in (\ref{eq:cubd}). If we then introduce a defect $h_\alpha$, the defect fixed points will follow from the analysis of hypercubic fixed points given in section \ref{hypercubic}. The defect couplings $h_i$ can then be obtained by
\begin{equation}
    h_i=(e_N)_i^\alpha h_\alpha\,.
\end{equation}

However, $T_N\neq C_{N+1}$, and there are two crucial features that distinguish between them. First, the bulk couplings remain as (\ref{eq:tetfixedpointnegative}) or (\ref{eq:tetfixedpointpositive}) even after we have transformed into this hypercubic-like form, and the $N$-dependence of the coefficient of $(\phi^2)^2$ in the $d$ tensor differs. This will only amount to altering the numerical values of the solutions, but will not affect the form of the solutions themselves. More importantly, the tetrahedral system has a constraint on its fields which is not present in the hypercubic model. From the first equation in (\ref{eq:evectorrules}), we see that the bulk field and the defect couplings ought to obey
\begin{equation}
    \sum_\alpha\phi^\alpha=0\,,\qquad\sum_\alpha h^\alpha=0\,.
\end{equation}
In terms of the hypercubic fixed points exhibited in section \ref{hypercubic}, which have all of the non-zero $h_\alpha$ equal to the same constant, this constraint is realised by restricting us to only consider fixed points at which an even number of the $h_\alpha$ are non-zero. Of these non-zero $h_\alpha$, half of them must be positive, with the other half negative. The $S_{N+1}\times\mathbb{Z}_2$ symmetry of $T_N$ then acts on these vectors by permuting their entries and multiplying by an overall minus sign. One can then easily see that the solutions will lie in a single orbit of $S_{N+1}\times\mathbb{Z}_2$. Hence, each $C_{N+1}$ solution with an even number of non-zero couplings will yield a single equivalence class of $T_N$ solutions.

When one looks at $\beta^\alpha=(e_N)_i^\alpha\beta_i$, one finds 
\begin{equation}\label{devecs}
    (e_N)_i^\alpha\lsp d_{ijkl}h_jh_kh_l=(h^\alpha)^3-\frac{1}{N+1}\sum_{\beta=1}^{N+1}(h^\beta)^3-\frac{3N}{(N+1)(N+2)}h^2h^\alpha \,.
\end{equation}
Were it not for the middle term, $T_N$ would be precisely equivalent to $C_{N+1}$ with the constraint $\sum_\alpha h^\alpha=0$. For solutions descending from hypercubic points this term vanishes, and one can show that it gives an additive positive-definite contribution to the stability matrix, so that these solutions inherit all of the properties from their $C_{N+1}$ antecedents one would naively expect. In general, however, this term may not vanish, leading to additional classes of solutions. For instance, one will have solutions in which $k$ of the $h_{\alpha}$ vanish, and the non-zero couplings take the form
\begin{equation}\label{solhansatz}
    h_\alpha=\frac{h}{m}\;\;\text{for $m$ $\alpha$}\,,\quad h_\alpha=-\frac{h}{N+1-k-m}\;\;\text{for the remaining $N+1-k-m$ $\alpha$}\,,
\end{equation}
for some $k\leq N+1$, $m\leq N+1$ such that $k+m\leq N+1$, where $h$ is then determined by the resulting beta functions. There are $2\binom{N+1}{k}\binom{N+1-k}{m}$ equivalent solutions in each class. A survey of solutions of $T_4$ reveal that there are also solutions which take a more complicated form, and it is likely that these isolated points proliferate both in number and complexity as $N$ increases. In the remainder of this subsection we report results for a few low values of $N$.

For $N=4$ at the $T_{4\lsp -}$ fixed point we find 81 solutions in 5 distinct universality classes. Among them there are 20 equivalent solutions that correspond to an IR stable dCFT. The eigenvalues of their corresponding stability matrix are $\varepsilon,\frac25\varepsilon,\frac{1}{10}\varepsilon(2)$, where we use the notation $x(y)$ with $x$ the eigenvalues of the stability matrix and $y$ their multiplicities. Since the order of the quotient of the symmetry breaking is 20, the order of the symmetry group preserved by the IR stable dCFT is $5!\times 2 / 20=12$. Among the subgroups of $S_5\times \mathbb{Z}_2$ with order 12 there is only one with a two-dimensional irreducible representation, namely the dihedral group $D_6$.\footnote{The other order-12 subgroups of $S_5\times\mathbb{Z}_2$ are the alternating group $A_4$ and $\mathbb{Z}_6\times\mathbb{Z}_2$.} The fact that there exists a multiplicity-two eigenvalue of the stability matrix of the IR stable dCFT then shows that its global symmetry group is $D_6$.

We find solutions among which only one corresponds to a stable fixed point for $N=5,6$, but for $N=7$ we have the first example of two distinct stable IR dCFTs. Here there are nine inequivalent non-trivial classes of solutions, two of which give IR stable fixed points. At the $\lambda_+,g_+$ fixed point we find the representative solutions
\begin{equation}\label{htfp1plus}
    h_{+, i}=\left(0,0,0,4 \sqrt{\tfrac{66}{85}},8 \sqrt{\tfrac{11}{85}},8 \sqrt{\tfrac{11}{119}},-3
   \sqrt{\tfrac{33}{119}}\right); \qquad \kappa=\{\varepsilon(1),\tfrac{14}{17}\varepsilon(2),\tfrac{2}{17}\varepsilon(4)\}\,;\qquad H=-\tfrac{495}{136}\varepsilon\,, 
\end{equation}
and
\begin{equation}\label{htfp2plus}
    h_{+, i}=\left(0,0,\tfrac32 \sqrt{\tfrac{11}{2}},\tfrac32 \sqrt{\tfrac{33}{10}},-\sqrt{\tfrac{11}{5}},2
   \sqrt{\tfrac{11}{7}},\sqrt{\tfrac{33}{7}}\right);\qquad \kappa=\{\varepsilon(1),\tfrac12\varepsilon(6)\}\,;\qquad H=-\tfrac{33}{8}\varepsilon\,.
\end{equation}
There is also a pair of IR stable fixed points for the $\lambda_-,g_-$ fixed point, represented by
\begin{equation}\label{htfp1minus}
    h_{-, i}=\left(0,-\sqrt{\tfrac{15}{2}},\tfrac{\sqrt{15}}{4},\tfrac34,\tfrac12\sqrt{\tfrac32},-\tfrac52\sqrt{\tfrac{15}{14}}, -\tfrac{15}{4}\sqrt{\tfrac{5}{14}}\right); \quad \kappa=\{\varepsilon(1),\tfrac{7}{16}\varepsilon(2),\tfrac{1}{16}\varepsilon(4)\}\,;\quad H=-\tfrac{675}{256}\varepsilon\,, 
\end{equation}
and
\begin{equation}\label{htfp2minus}
    h_{-, i}=\left(-\tfrac32\sqrt{\tfrac52},-\tfrac12\sqrt{\tfrac{15}{2}},-\tfrac{\sqrt{15}}{4},\tfrac94,\tfrac32\sqrt{\tfrac32},-\tfrac32\sqrt{\tfrac{15}{14}},3\sqrt{\tfrac{5}{14}}\right);\qquad \kappa=\{\varepsilon(1),\tfrac14\varepsilon(6)\}\,;\qquad H=-\tfrac{45}{16}\varepsilon\,.
\end{equation}
The fact that the stability matrices have different multiplicity eigenvalues in \eqref{htfp1plus} and \eqref{htfp2plus} (as well as in \eqref{htfp1minus} and \eqref{htfp2minus}) shows that these dCFTs have different symmetry. While we have not been able to determine the global symmetry groups of these dCFTs, we have found that the solutions \eqref{htfp2plus} and \eqref{htfp2minus} descend from the $N=8$ hypercubic case as described above, and the order of their symmetry group is 1152. The solutions \eqref{htfp1plus} and \eqref{htfp1minus} arise from setting $k=0,m=3$ in \eqref{solhansatz}, and their symmetry group has order 720. We encounter a similar situation of two distinct stable IR dCFTs for $N=9$ in our explicit calculations.

The solutions arising as hypercubic fixed points allow us to understand the origin of these multiple stable fixed points. The stable fixed point for a hypercubic model is, from section \ref{hypercubic}, the one in which $h_\alpha^2$ are all non-zero and equal. For $T_N$, the $C_{N+1}$ stable fixed point will only be consistent with the constraint $\sum_\alpha h^\alpha=0$ for $N$ odd. As this point continues to be stable as a $T_N$ solution, $T_N$ for $N$ odd will necessarily see at least one stable fixed point arising from a hypercubic solution. Importantly, it seems that the class of solutions with $\sum_\alpha (h^\alpha)^3\neq0$ is also able to independently provide stable solutions. The $N=4$ stable fixed point must, and indeed does, arise from this second class, just as one of the $N=7,9$ stable fixed points comes from the stable $N=8,10$ hypercubic solution, while the other lies in the second class. It seems likely that this pattern will continue, and that for larger odd $N$ theories will again have multiple stable defect fixed points.

\subsection{\texorpdfstring{$O(m)\times O(n)$}{O(m)xO(n)} biconical model}
The $O(m)\times O(n)$ biconical model has two quadratic invariants and three symmetric traceless rank four invariant tensors, so it goes beyond the class of examples described by \eqref{lag}. Nevertheless, it can be treated in a similar way---see \cite[Appendix B]{Osborn:2017ucf}. The action is
\begin{equation}
    S=\int d^{\lsp d}x\,\big(\tfrac12\partial^\mu\hat{\phi}_{\hat{\imath}}\lsp \partial^\mu\hat{\phi}_{\hat{\imath}}+\tfrac12\partial^\mu\check{\phi}_{\check{\imath}}\lsp \partial^\mu\check{\phi}_{\check{\imath}}+\tfrac18\lambda_1(\hat{\phi}^2)^2+\tfrac18\lambda_2(\check{\phi}^2)^2+\tfrac14 g\hat{\phi}^2\check{\phi}^2\big)\,,
\end{equation}
where $\hat{\phi}_{\hat{\imath}}$, $\hat{\imath}=1,\ldots,m$ are the fields transforming under $O(m)$ and $\check{\phi}_{\check{\imath}}$, $\check{\imath}=1,\ldots,n$ the fields transforming under $O(n)$.
The location of the biconical fixed point in the space of $\lambda_1,\lambda_2,g$ couplings is a complicated function of $m$ and $n$, which simplifies considerably when $m=n$:
\begin{equation}
    \lambda_1=\lambda_2=\frac{n}{2(n^2+8)}\varepsilon\,,\qquad g=-\frac{n-4}{2(n^2+8)}\varepsilon\qquad\qquad (m=n)\,.
\end{equation}
This is the stable fixed point for $2<n<4$. For $m=n=4$ it coincides with two decoupled $O(4)$ models, but at higher orders in $\varepsilon$ the $n$ for which the $m=n$ biconical theory coincides with the decoupled one receives corrections~\cite[Appendix B]{Osborn:2017ucf}.

In the biconical model we may discuss the defect deformation
\begin{equation}
    S\to S'=S+\hat{h}_{\hat{\imath}}\int d\tau\,\hat{\phi}_{\hat{\imath}}+\check{h}_{\check{\imath}}\int d\tau\,\check{\phi}_{\check{\imath}}\,.
\end{equation}
We find
\begin{equation}
    \hat{\beta}_{\hat{\imath}}=-\tfrac12 \hat{h}_{\hat{\imath}}(\varepsilon-\lambda_1 \hat{h}^2-g\check{h}^2)\,,\qquad
    \check{\beta}_{\check{\imath}}=-\tfrac12 \check{h}_{\check{\imath}}(\varepsilon-\lambda_2 \check{h}^2-g\hat{h}^2)\,,
\end{equation}
with non-trivial roots occurring at
\begin{equation}\label{defbiconfi}
    \hat{h}^2=\frac{g-\lambda_2}{g^2-\lambda_1\lambda_2}\varepsilon\,,\qquad \check{h}^2=\frac{g-\lambda_1}{g^2-\lambda_1\lambda_2}\varepsilon\,,
\end{equation}
or
\begin{equation}
    \hat{h}^2=\frac{1}{\lambda_1}\varepsilon\,,\qquad \check{h}_{\check{\imath}}=0\,,
\end{equation}
or
\begin{equation}
    \hat{h}_{\hat{\imath}}=0\,,\qquad \check{h}^2=\frac{1}{\lambda_2}\varepsilon\,.
\end{equation}
The preserved symmetry is $O(m-1)\times O(n-1)$, or $O(m-1)\times O(n)$ or $O(m)\times O(n-1)$, respectively.

For $m=n$ we correspondingly find
\begin{equation}
    \hat{h}^2=\check{h}^2=\tfrac12(n^2+8)\,,\qquad \text{or}\qquad \hat{h}^2=\frac{2}{n}(n^2+8)\,,\,\check{h}_{\check{\imath}}=0\,,\qquad\text{or}\qquad \hat{h}_{\hat{\imath}}=0\,,\,\check{h}^2=\frac{2}{n}(n^2+8)\,.
\end{equation}

The analysis of stability proceeds in a straightforward manner. Corresponding to the different solutions above we find for the stability matrix that
\begin{equation}
   S_{IJ}=\begin{pmatrix} \lambda_1 \hat{h}_{\hat{\imath}}\hat{h}_{\hat{\jmath}} && g\hat{h}_{\hat{\imath}}\check{h}_{\check{\jmath}} \\ g\check{h}_{\check{\imath}}\hat{h}_{\hat{\jmath}} && \lambda_2\check{h}_{\check{\imath}}\check{h}_{\check{\jmath}}
   \end{pmatrix}\,,
   \label{bothhnonzero}
\end{equation}
or
\begin{equation}
    S_{IJ}=
    \begin{pmatrix}\lambda_1 \hat{h}_{\hat{\imath}}\hat{h}_{\hat{\jmath}} && 0 \\ 0 && \frac{\varepsilon}{2}\big(\frac{g}{\lambda_1}-1\big)\delta_{\hat{\imath}\hat{\jmath}}
    \end{pmatrix}\,,
    \label{hcheckzero}
\end{equation}
or
\begin{equation}
    S_{IJ}=
    \begin{pmatrix}
    \frac{\varepsilon}{2}\big(\frac{g}{\lambda_2}-1\big)\delta_{\hat{\imath}\hat{\jmath}} && 0 \\ 0 && \lambda_2\check{h}_{\check{\imath}}\check{h}_{\check{\jmath}}
    \end{pmatrix}\,.
    \label{hhatzero}
\end{equation}
Noting that $\lambda_1,\lambda_2>g$,\foot{Positivity of the scalar potential of the bulk biconical model requires $\lambda_1,\lambda_2>0$ and $\lambda_1\lambda_2>g^2$. Unitarity of the defect CFT as defined by \eqref{defbiconfi} then requires $\lambda_1,\lambda_2>g$.} we see that both (\ref{hcheckzero}) and (\ref{hhatzero}) have negative eigenvalues. Thus, the fixed points with one of the defects being trivial will be unstable. To see that the fixed point (\ref{bothhnonzero}) is in fact stable, we note that there will be $m-1$ vectors $v_{\hat{\imath}}$ orthogonal to $\hat{h}_{\hat{\imath}}$ and $n-1$ vectors $u_{\check{\imath}}$ orthogonal to $\check{h}_{\check{\imath}}$, giving us $m+n-2$ vectors in the kernel of the stability matrix:
\begin{equation}
    S\begin{pmatrix}v_{\hat{\jmath}}\\0\end{pmatrix}=0\,,\qquad S\begin{pmatrix}0\\u_{\check{\jmath}}\end{pmatrix}=0\,.
\end{equation}
To find the remaining two eigenvectors, we note that the beta functions require that the solution itself must be an eigenvector of the stability matrix with eigenvalue $\varepsilon$, that is
\begin{equation}
    S\begin{pmatrix}\hat{h}_{\hat{\jmath}}\\\check{h}_{\check{\jmath}}\end{pmatrix}=\begin{pmatrix}\hat{h}_{\hat{\imath}}(\lambda_1\hat{h}^2+g\check{h}^2)\\\check{h}_{\check{\imath}}(g\hat{h}^2+\lambda_2\check{h}^2)\end{pmatrix}=\varepsilon\begin{pmatrix}\hat{h}_{\hat{\imath}}\\\check{h}_{\check{\imath}}\end{pmatrix}\,.
\end{equation}
The trace of the stability matrix can easily be seen to be
\begin{equation}
    \Tr S=\lambda_1\hat{h}^2+\lambda_2\check{h}^2=\varepsilon+\frac{(g-\lambda_1)(g-\lambda_2)}{\lambda_1\lambda_2-g^2}\varepsilon\,,
\end{equation}
so that the last eigenvalue will be
\begin{equation}
    \kappa=\frac{(g-\lambda_1)(g-\lambda_2)}{\lambda_1\lambda_2-g^2}\varepsilon>0\,.
\end{equation}
The corresponding eigenvector can also be determined:
\begin{equation}
    S\begin{pmatrix}
        -\sqrt{\frac{g-\lambda_1}{g-\lambda_2}}\hat{h}_{\hat{\jmath}} \\ \sqrt{\frac{g-\lambda_2}{g-\lambda_1}}\check{h}_{\check{\jmath}}
    \end{pmatrix}=\frac{(g-\lambda_1)(g-\lambda_2)}{\lambda_1\lambda_2-g^2}\varepsilon\begin{pmatrix}
        -\sqrt{\frac{g-\lambda_1}{g-\lambda_2}}\hat{h}_{\hat{\jmath}} \\ \sqrt{\frac{g-\lambda_2}{g-\lambda_1}}\check{h}_{\check{\jmath}}
    \end{pmatrix}\,.
    \label{othereigenvector}
\end{equation}
As $S_{IJ}$ has no negative eigenvalues, the dCFT described by (\ref{defbiconfi}) will be stable. 

Loosening the restrictions of $\lambda_1,\,\lambda_2,$ and $g$ brought by unitarity, the stability of the fixed points can shift. If one relaxes this requirement, thus allowing $\lambda_1<g$ or $\lambda_2<g$, one can arrive at different stability configurations. Positivity of the potential in the bulk prevents $\lambda_1,\,\lambda_2<g$ simultaneously, so that one finds only two different scenarios. In both cases (\ref{bothhnonzero}) will become unstable due to the eigenvalue $\kappa$ becoming negative. If $\lambda_1<g<\lambda_2$ then (\ref{hcheckzero}) will become the stable fixed point, while if $\lambda_1<g<\lambda_2$ (\ref{hhatzero}) become stable.

\subsection{MN model}
The MN model refers to CFTs with global symmetry $O(m)^n\rtimes S_n$. It has $N=mn$ scalars, and one way to describe it is by decomposing $\phi_i$ into $n$ vectors $\vec{\varphi}_r$ of size $m$ each. Then, it can be written in the form of \eqref{lag} with
\begin{equation}
    d_{ijkl}\phi_i\phi_j\phi_k\phi_l=\sum_r (\vec{\varphi}_r^{\,2})^2-\frac{m+2}{N+2}(\vec{\varphi}^{\,2})^2\,,\qquad \vec{\varphi}^{\,2}=\sum_r\vec{\varphi}_r^{\,2}\,.
\end{equation}
There are two fixed points of this type, with one being that of $n$ decoupled $O(m)$ theories and the other fully interacting. The fully interacting fixed point is RG stable for $m<4$ and has
\begin{equation}
    \lambda=\frac{6(N-m)}{(N+2)((m+8)N-16(m-1))}\varepsilon\,,\qquad g=\frac{3(N-4)}{(m+8)N-16(m-1)}\varepsilon\,.
\end{equation}
The other fixed point in \eqref{fpcoup} corresponds to $n$ decoupled $O(m)$ models. For $m=1$ the fully interacting MN fixed point reduces to the hypercubic one. For $m=4$ we have reduction to the case of $n$ decoupled $O(4)$ models. For $m=n=2$ we have coincidence with the $O(4)$ model.

With the defect, we will find it useful to put the interaction action in a form reminiscent of the $O(m)\times O(n)$ biconical case. To this end, we will not use $\lambda$ and $g$, but instead the combination
\begin{equation}
   u=\tfrac{1}{3}\Big(3\lambda+\frac{N-m}{N+2}g\Big)\,,\qquad v=\tfrac{2}{3}\Big(3\lambda-\frac{m+2}{N+2}g\Big)\,,
\end{equation}
so that the scalar interaction takes the more convenient form
\begin{equation}
\lambda_{ijkl}\phi_i\phi_j\phi_k\phi_l=3\bigg(u\sum_{r}(\vec{\varphi}_{r}^{\,2})^2+v\sum_{r<s}\vec{\varphi}_{r}^{\,2}\vec{\varphi}_{s}^{\,2}\bigg)\,.
\end{equation}
Focusing on the fully interacting fixed point, we introduce the defect
\begin{equation}
    \vec{h}_r\cdot\int d\tau\, \vec{\varphi}_r(\tau,\mathbf{0})=h_r^a\int d\tau\, \varphi_r^a(\tau,\mathbf{0})\,,\qquad a=1,\ldots,m\,.
\end{equation}
The defect couplings will have the beta functions
\begin{equation}
    \beta_r^a=-\tfrac{1}{2}h_r^a
    \Big(\varepsilon-u \vec{h}_r^2-v\sum_{s\neq r}\vec{h}_s^2\Big)\,.
\end{equation}
The fixed points of this model take a very similar form to those of the $O(m)\times O(n)$ biconical model, and can then be divided into $n$ classes depending on the number $k$ of trivial defect coupling vectors, where $0\leq k\leq n$. Using the $S_n$ symmetry we can choose the trivial coupling vectors to be the first $k$, with the other $n-k$ being equal in magnitude and satisfying (for $k\neq n$)
\begin{equation}
    \vec{h}_r^2=\frac{\varepsilon}{u+(n-k-1)v}\,,\qquad r=k+1,\ldots,n\,.
\end{equation}

The fixed point will not be stable for $k\neq0$. To see this, notice that the defect beta function has derivatives
\begin{equation}
    \begin{aligned}
        \partial_r^a \beta_r^b&=-\tfrac{1}{2}\delta^{ab}\Big(\varepsilon-u \vec{h}_r^2-v\sum_{r\neq r}\vec{h}_s^2\Big)+u h_r^a h_r^b\quad\quad\quad\text{(no sum on }r)\,, \\
        \partial_s^b\beta_r^a&=vh_r^a h_s^b\quad\quad\quad (r\neq s)\,.
    \end{aligned}
\end{equation}
Notice that if $h_r^a=0$, then $\partial_r^a \beta_r^b\neq0$. For $k\neq0$ we can permute the indices to choose $\vec{h}_1=0$, so that the stability matrix will take the form
\begin{equation}
    S_{IJ}=\begin{pmatrix} \frac{\varepsilon}{2}\Big(\frac{(n-k)v}{(n-k)v-v+u}-1\Big)\delta^{ab} && 0 \\ 0 && S'_{IJ}
    \end{pmatrix}\,,
\end{equation}
where $S'_{IJ}$ is the rest of the stability matrix. As with the biconical model, positivity and unitarity demand that $u>0$ and $v<u$, so that the upper-left block will be a diagonal matrix with negative elements. Thus, $S_{IJ}$ will have at least $m$ negative eigenvalues. The symmetry is broken at these fixed points to $\big( O(m)^k\rtimes S_k\big)\times\big(O(m-1)^{n-k}\rtimes S_{n-k}\big)$.

For the $k=0$ fixed point, where $\vec{h}_r^2=h^2=\frac{\varepsilon}{u+(n-1)v}$ for all $r=1,\ldots,n$, the stability matrix takes the form
\begin{equation}
    S_{IJ}=\begin{pmatrix}
        u h_1^a h_1^b && vh_1^a h_2^b &&\cdots \\ vh_2^a h_1^b && u h_2^a h_2^b &&\cdots\\
        \vdots && \vdots && \ddots
    \end{pmatrix}\,.
\end{equation}
This is very similar to Eq.~(\ref{bothhnonzero}), and the analysis proceeds similarly. For each $r$, there will be $m-1$ vectors in the kernel corresponding to vectors orthogonal to $h_r^a$. There will be a single eigenvector with eigenvalue $\varepsilon$ corresponding to the perturbation itself:
\begin{equation}
    S\begin{pmatrix}
        h_1^a \\ h_2^a \\ \vdots
    \end{pmatrix}=(u+(n-1)v)h^2\begin{pmatrix}
         h_1^a \\ h_2^a \\ \vdots
    \end{pmatrix}=\varepsilon\begin{pmatrix}
         h_1^a \\ h_2^a \\ \vdots
    \end{pmatrix}\,.
\end{equation}
The remaining $n-1$ eigenvectors have eigenvalue $\kappa=(u-v)h^2>0$ and are essentially generalisations of Eq.~(\ref{othereigenvector}), with only two elements being non-zero in each,
\begin{equation}
   S\begin{pmatrix}
        -h_1^a\\0\\\vdots\\0\\h_r^a\\0\\\vdots
    \end{pmatrix}=(u-v)h^2\begin{pmatrix}
        -h_1^a\\0\\\vdots\\0\\h_r^a\\0\\\vdots
    \end{pmatrix}\qquad (r=2,\ldots, n)\,.
\end{equation}
As $S_{IJ}$ has no negative eigenvalues, this fixed point will be stable. Its symmetry is $O(m-1)^n\rtimes S_n$.

\section{Adding fermions}\label{fermionsec}
With fermions in the bulk we consider the action
\eqref{scferLag}. At any non-trivial scalar-fermion fixed point, the dimension of the $\phi_i$'s will be given by the eigenvalues of the matrix
\begin{equation}
    (\Delta_{\phi})_{ij}=(1-\tfrac12\varepsilon)\delta_{ij}+\tfrac12 Y_{ij}\,.
\end{equation}
Eigenvalues that are below 1 will correspond to scalar fields that can serve as non-trivial line defect deformations. Assuming that that is indeed the case for $n_s$ of the $N_s$ scalars, we can consider deformations of the form \eqref{defdeform}. Due to the presence of fermions there is now an extra term in $\beta_i$ compared to \eqref{betahgen}. As we saw above, at leading order in the bulk couplings we have
\begin{equation}\label{betahgenfer}
    \beta_i=-\tfrac12\varepsilon\lsp h_i+\tfrac16\lsp \lambda_{ijkl}h_jh_kh_l+\tfrac12 Y_{ij}h_j\,.
\end{equation}
The RG flow is again gradient with
\begin{equation}
    H=-\tfrac14\varepsilon\lsp h^2+\tfrac{1}{24}\lsp\lambda_{ijkl}h_ih_jh_kh_l+\tfrac14 Y_{ij}h_ih_j\,.
\end{equation}

\subsection{Gross--Neveu--Yukawa model}
This Gross--Neveu--Yukawa (GNY) was discussed in detail in \cite{Giombi:2022vnz}. The action is
\begin{equation}\label{lagGNY}
    S_{\text{GNY}}= \int d^{\lsp d}x\,\big(\tfrac12\partial^\mu\phi\lsp\partial_\mu\phi + i\lsp \overbar{\Psi}_a\slashed{\partial}\Psi_a+ y\lsp\phi\overbar{\Psi}_a\Psi_a +\tfrac18\lambda \phi^4\big)\,,
\end{equation}
with one real scalar $\phi$ and $N_f$ Dirac fermions $\Psi_a, a=1,\ldots,N_f$. We define $\overbar{\Psi}=\Psi^\dagger\gamma^0$. In our conventions the beta functions of $y$ and $\lambda$ at leading order are
\begin{equation}
    \beta_y=-\tfrac12\varepsilon\lsp y+\tfrac12(N+6)y^3\,,\qquad \beta_\lambda=-\varepsilon\lsp\lambda+9\lsp\lambda^2+2N\lsp\lambda y^2-4N\lsp y^4\,,
\end{equation}
where $N=4N_f$. A fixed point occurs for
\begin{equation}
    y^2=\frac{1}{N+6}\varepsilon\,,\qquad \lambda=\frac{\sqrt{P_N}-N+6}{18(N+6)}\varepsilon\,,
\end{equation}
where $P_N=N^2+132N+36$. At that fixed point we have
\begin{equation}
    \Delta_\phi=1-\frac{3}{N+6}\varepsilon<1\,,
\end{equation}
and we may consider the defect deformation
\begin{equation}
    S_{\text{GNY}}\to S'_{\text{GNY}}=S_{\text{GNY}}+h\int d\tau\,\phi(\tau,\mathbf{0})\,.
\end{equation}
For the defect coupling we have\footnote{Here $Y_{ij}\to Ny^2$ and $\widetilde{Y}_{ijkl}\to Ny^4$.}
\begin{equation}
    \beta_h=-\tfrac12h(\varepsilon-\lambda h^2-N\lsp y^2)\,,
    \label{gnydefectbeta}
\end{equation}
and a non-trivial fixed point is found for
\begin{equation}
    h^2=\frac{108}{\sqrt{P_N}-N+6}\,.
\end{equation}
This is a stable fixed point, since $\Delta_\phi=1+\frac{6}{N+6}\varepsilon>1$.

Let us remark here that the bulk GNY model has emergent supersymmetry when $N=1$, which requires a fractional number of Dirac fermions in \eqref{lagGNY}, namely $N_f=\frac14$. With this choice and in the limit $\varepsilon\to1$ the GNY model would have one 3D Majorana spinor and two supercharges~\cite{Fei:2016sgs}.

\subsection{Nambu--Jona-Lasinio--Yukawa model}
The Nambu--Jona-Lasinio--Yukawa (NJLY) model has two real scalar fields $\phi_1$ and $\phi_2$ ($\phi_2$ is a pseudoscalar) and $N_f$ Dirac fermions $\Psi_a, a=1,\ldots,N_f$. It is sometimes called the chiral XY model; see e.g.\ \cite{Zerf:2017zqi}. Its action is
\begin{equation}\label{lagNJLY}
    S_{\text{NJLY}}= \int d^{\lsp d}x\,\big(\tfrac12\partial^\mu\phi_1\lsp\partial_\mu\phi_1 + \tfrac12\partial^\mu\phi_2\lsp\partial_\mu\phi_2 + i\lsp \overbar{\Psi}_a\slashed{\partial}\Psi_a + y\lsp\overbar{\Psi}_a(\phi_1+i\lsp\gamma^5\phi_2)\Psi_a+\tfrac{1}{8}\lambda(\phi_1^2+\phi_2^2)^2\big)\,,
\end{equation}
and it has a chiral $U(1)$ symmetry generated by
\begin{equation}
    \phi=\phi_1+i\phi_2\to e^{-2i\alpha}\phi\,,\qquad \Psi_a\to e^{i\alpha\gamma^5}\Psi_a\,.
\end{equation}
The beta functions of $y$ and $\lambda$ at leading order are
\begin{equation}
    \beta_y=-\tfrac12\varepsilon\lsp y+\tfrac12(N+4)y^3\,,\qquad \beta_\lambda=-\varepsilon\lsp\lambda+10\lsp \lambda^2+2N\lambda y^2-4Ny^4\,,
\end{equation}
where $N=4N_f$, and the NJLY fixed point lies at
\begin{equation}
    y^2=\frac{1}{N+4}\varepsilon\,,\qquad \lambda=\frac{\sqrt{R_N}-N+4}{20(N+4)}\varepsilon\,,
\end{equation}
where $R_N=N^2+152\lsp N+16$.

At this fixed point one may compute
\begin{equation}
    \Delta_{\phi_1}=\Delta_{\phi_2}=1-\frac{2}{N+4}\varepsilon\,,
\end{equation}
and since $\Delta_{\phi_i}<1$ one may discuss the line defect deformation
\begin{equation}
    S_{\text{NJLY}}\to S'_{\text{NJLY}} = S_{\text{NJLY}} + h_1\int d\tau\,\phi_1(\tau,\mathbf{0}) + h_2\int d\tau\,\phi_2(\tau,\mathbf{0})\,.
\end{equation}
It is straightforward to compute\footnote{Here $Y_{ij}=Ny^2\delta_{ij}$ and $\widetilde{Y}_{ijkl}=Ny^4(\delta_{ij}\delta_{kl}-\delta_{ik}\delta_{jl}+\delta_{il}\delta_{jk})$.}
\begin{equation}
    \beta_i=-\tfrac12\lsp h_i(\varepsilon-\lambda h^2-Ny^2)\,,\qquad i=1,2\,,\qquad h^2=h_1^2+h_2^2\,,
    \label{njlydefectbeta}
\end{equation}
and a non-trivial root is found for
\begin{equation}\label{hNJLY}
    h^2=\frac{80}{\sqrt{R_N}-N+4}\,.
\end{equation}
Due to the $U(1)$ symmetry of the NJLY model we see that we can fix $h_1^2+h_2^2$ but not $h_1$ and $h_2$ separately. Obviously the line defect breaks the $U(1)$ symmetry completely.

Looking at the stability matrix of the IR dCFT defined by \eqref{hNJLY} we find one irrelevant operator given by $\mathcal{O}=h_i\phi_i$ with dimension $\Delta_{\mathcal{O}}=1+\frac{4}{N+4}\varepsilon$, and the marginal (tilt) operator $\mathcal{O}'=h_1\phi_2-h_2\phi_1$.

The NJLY model has emergent supersymmetry when $N=2$, which requires a fractional number of Dirac fermions in \eqref{lagNJLY}, namely $N_f=\frac12$. With this choice and in the limit $\varepsilon\to1$ the NJLY model would have one 3D Dirac spinor and four supercharges~\cite{Fei:2016sgs}.

\subsection{Chiral Heisenberg model}
The chiral Heisenberg (cH) model is a generalisation of the GNY model useful for modelling the semimetallic-antiferromagnetic phase transition in graphene~\cite{Zerf:2017zqi, Janssen:2014gea, Rosenstein:1993zf}. The theory is governed by the action
\begin{equation}
    S_{\text{cH}}= \int d^{\lsp d}x\,\big(\tfrac12\partial^\mu\phi_i\lsp\partial_\mu\phi_i+i\overbar{\Psi}(\mathds{1}_{2}\otimes\gamma^\mu)\partial_\mu\Psi+y\lsp\phi_i\overbar{\Psi}\big(\sigma_i\otimes\mathds{1}_{2N_f}\big)\Psi+\tfrac18\lambda(\phi^2)^2\big)\,,
\end{equation}
containing three real scalar fields $\phi_i$, $i=1,2,3$, $N_f$ Dirac fermions arranged into two larger spinors
\begin{equation}
    \Psi=\begin{pmatrix}
        \Psi_{+} \\ \Psi_{-}
    \end{pmatrix}\,, \qquad \overbar{\Psi}=\Psi^\dagger(\mathds{1}_{2}\otimes\gamma^0)\,,
\end{equation}
and where $\sigma_i$ are the Pauli matrices. For $N_f=2$ the spinors $\Psi_{\pm}$ will be the usual four-component Dirac spinors, but at the level of the beta function one can imagine taking $N_f$ to be a continuous parameter, with each $\Psi_{\pm}$ then containing $2\lsp N_f$ complex components\cite{Zerf:2017zqi}. This action retains an $SO(3)$ symmetry associated with the rotations,
\begin{equation}
    \phi_i\rightarrow R_{ij}\phi_j\,,\qquad \Psi\rightarrow e^{i\theta\lsp \Vec{n}\cdot(\Vec{\sigma}\otimes \mathds{1}_{2N_f})}\Psi\,,
\end{equation}
where $R_{ij}$ is the $SO(3)$ matrix associated with a rotation about $\Vec{n}$ by an angle $\theta$. If we again use $N=4N_f$, the beta functions for the couplings are given to one-loop order by \cite{Rosenstein:1993zf}\foot{This example does not belong to the class of examples captured by \eqref{scferLag}. Nevertheless, the one-loop beta function for the defect depends only on the fermionic coupling through $\gamma_\phi$, and thus takes a similar form to \eqref{gnydefectbeta} and \eqref{njlydefectbeta}.}
\begin{equation}
    \beta_y=-\tfrac12\varepsilon\lsp y+\tfrac12(N+2)y^3\,,\qquad\beta_\lambda=-\varepsilon\lambda+11\lambda^2+2N\lambda y^2-4Ny^4\,,
\end{equation}
from which one finds the non-trivial fixed point
\begin{equation}
    y^2=\frac{1}{N+2}\varepsilon\,, \qquad \lambda=\frac{\sqrt{S_N}-N+2}{22(N+2)}\varepsilon\,,
\end{equation}
where $S_N=N^2+172N+4$. At the fixed point, the scaling dimension of the scalar fields is
\begin{equation}
    \Delta_{\phi_i}=1-\frac{1}{N+2}\varepsilon<1\,,
\end{equation}
so that we can add a relevant defect deformation:
\begin{equation}
    S_{\text{cH}}\to S_{\text{cH}}'=S_{\text{cH}}+h_i\int d\tau\,\phi_i(\tau,\mathbf{0})\,.
\end{equation}

To leading order we find the beta function for the defect coupling to be
\begin{equation}\label{defspherecH}
    \beta_i=-\tfrac12h_i(\varepsilon-\lambda h^2-Ny^2)\,.
\end{equation}
Besides the trivial $h_i=0$ solution, at the chiral Heisenberg point there is the additional solution
\begin{equation}
    h^2=\frac{44}{\sqrt{S_N}-N+2}\,.
\end{equation}
Much as with the GNY and NJLY models, analysing the stability matrix at this point shows that again we have one irrelevant operator $\mathcal{O}=h_i\phi_i$ with dimension $\Delta_{\mathcal{O}}=1+\frac{2}{N+2}\varepsilon$, and two marginal operators $\mathcal{O}_1=h_1\phi_2-h_2\phi_1$ and $\mathcal{O}_2=h_1\phi_3-h_3\phi_1$. The presence of the two marginal operators is related to the breaking of the bulk $SO(3)$ symmetry to $SO(2)$ on the defect. The associated quotient is $SO(3)/SO(2)$, which is isomorphic to the two-sphere $S^2$.

\section{Conclusion}\label{conc}
Beginning with a general action for both a scalar and a scalar-fermion system, we have found the beta function for a scalar line defect coupling to next-to-leading order in the bulk parameters. Using this general form, we have explored defects in a number of different scalar and scalar-fermion theories. Importantly, we note that, unlike in the bulk scalar system, the uniqueness of the stable defect fixed point is not guaranteed. However, we have only found one example, the hypertetrahedral model, where uniqueness is not observed, perhaps indicating that multiple stable fixed points requires very specific interactions. As stability can be seen to depend on the size of the defect coupling, $h^2$, this is likely due to the symmetry of the other systems we have considered greatly restricting the number and form of defect fixed points.

One could quite simply continue with this programme and examine defects in further scalar or scalar-fermion theories. For example, one could consider a defect in a scalar bulk theory with $O(m)\times O(n)$ or $U(m)\times U(n)$ symmetry (see e.g. \cite{Osborn:2017ucf,Kousvos:2022ewl}), however our brief preliminary examination of these theories do not indicate either a nice analytic form for the defect fixed point, or any interesting behaviour such as multiple stable fixed points. A more general brute-force numerical search, in the vein of \cite{Osborn:2020cnf}, may be able to reveal more unusual bulk theories which can be associated with multiple stable defect fixed points. One can also pursue studies of our theories with numerical bootstrap methods, as was done for the $O(2)$ line defect in~\cite{Gimenez-Grau:2022czc}, although in cases with discrete symmetries in the bulk there are no tilt operators. Computations of higher-point functions and analytic bootstrap studies could also be performed along the lines of~\cite{Lemos:2017vnx, Gimenez-Grau:2022ebb, Bianchi:2022sbz}.

The patterns of symmetry breaking may be understood by inspection of the defect beta functions, at least in simple cases. It is obvious that for the bulk $O(N)$ model the defect CFT will have $O(N-1)$ symmetry, but other cases appear to be more complicated. For example, in the cubic case one can find dCFTs where the bulk $\mathbb{Z}_2{\!}^3\rtimes S_3$ symmetry is broken to $D_4, \mathbb{Z}_2{\!}^2$ or $S_3$, see Fig.~\ref{cubicbr}, presumably due to the fact that the breaking is due to a deformation proportional to $\phi_i$, which transforms in the defining representation of the bulk global symmetry. It would be beneficial to develop general diagnostics for the patterns of symmetry breaking that may be obtained with line defect deformations of CFTs in the $\varepsilon$ expansion.

As far as applications go, it is well-known that the cubic and Heisenberg models in three dimensions are very hard to distinguish experimentally in $d=3$ due to the fact that their most easily accessible critical exponents are nearly identical. As we have seen in this work, the presence of a pinning field in these cases will have very different consequences: the $O(3)$ symmetry of the Heisenberg model will be broken to $O(2)$, while the $\mathbb{Z}_2{\!}^3\rtimes S_3$ symmetry of the cubic model will be broken to $D_4$ in the corresponding IR dCFTs. Potential experimental consequences of this may be of relevance in determining the universality class of systems like cubic magnets at criticality without relying on measurements of critical exponents. 

The one-point function of the order parameter in the presence of the defect has coefficients
\begin{equation}
    a_\phi^2=\tfrac{11}{4}+\tfrac14(11\log 2-1)\varepsilon\quad (\text{Heisenberg})\,,\qquad
    a_\phi^2=\tfrac{27}{8}+ \tfrac18(27\log 2-\tfrac{179}{18})\varepsilon\quad(\text{cubic})
\end{equation}
at next-to-leading order in the corresponding IR stable dCFTs.\footnote{We define $a_\phi$ via $\mu^{\varepsilon/2}\langle\phi(0,\mathbf{x})\rangle=a_\phi\sqrt{\Gamma(d/2-1)}/2\pi^{d/4}|\mathbf{x}|$. Note that the sign of $a_\phi$ does not have a physical meaning.} These evaluate to approximately 4.406 for Heisenberg and 4.471 for cubic if we use $\varepsilon=1$. As we observe, the order-$\varepsilon$ correction reduces the difference in these coefficients obtained from the leading term by an order of magnitude. Higher order corrections can be computed and a more trustworthy estimate of $a_\phi^2$ can then be made in the $\varepsilon\to 1$ limit in these theories. A sufficiently different one-point function coefficient between the Heisenberg and cubic cases could be useful in distinguishing these universality classes in experiments.

\ack{We thank C.\ Herzog, H.\ Osborn, M.\ Preti, M.\ Tr\'epanier and especially N.\ Drukker for enlightening discussions and comments on the manuscript. We have benefited from the use of GAP~\cite{GAP4}. AS is funded by the Royal Society under grant URF{\textbackslash}R1{\textbackslash}211417.}

\begin{appendices}

\section{Contributions to defect coupling beta function from fermions in the bulk}\label{app:fermions}

In this appendix, we exhibit the derivation of the defect counterterms leading to Eq.\ (\ref{BetaDefScalarFermion}). With the inclusion of fermions, our bulk action takes the form
\begin{equation}
    S=\int d^{\lsp d}x\,\big(\tfrac12\lsp\partial^\mu\phi_i\lsp\partial_\mu\phi_i+i\bar{\psi}_a\bar{\sigma}^\mu\partial_\mu\psi_a+\tfrac{1}{4!}\lambda_{ijkl}\lsp\phi_i\phi_j\phi_k\phi_l+(\tfrac12\lsp  y_{iab}\lsp\phi_i\psi_a\psi_b+\text{h.c.})\big)\,,
\end{equation}
so that the addition of the defect brings our action to
\begin{equation}
    S'=S+h_i\int d\tau\, \phi_i(\tau,\mathbf{0})\,.
\end{equation}
In addition to Eq.\ (\ref{scalarrules}), the presence of fermions gives us the additional rules
\begin{equation}
    \begin{aligned}
    \begin{tikzpicture}[baseline=(vert_cent.base)]
        \node (vert_cent) at (0,0) {$\phantom{\cdot}$};
        \draw[solid] node[at start,xshift=-14pt,yshift=0.5pt] {$x_1,\dot{\alpha}$} (0,0)--(1,0) node[at end,xshift=15pt,yshift=-1pt] {$x_2,\alpha$};
    \end{tikzpicture} &= \frac{(d-2)\Gamma(\tfrac12d-1)}{4\pi^{d/2}}\frac{x_{12}^{\;\;\; \mu}\bar{\sigma}_\mu^{\dot{\alpha}\alpha}}{(x_{12}^{\;\;\; 2})^{d/2}}\,,\\
    \begin{tikzpicture}[baseline=(vert_cent.base)]
        \node (vert_cent) at (0,0) {$\phantom{\cdot}$};
        \draw[solid] node[at start,xshift=-14pt,yshift=-1pt] {$x_1,\alpha$} (0,0)--(1,0) node[at end,xshift=15pt,yshift=0.5pt] {$x_2,\dot{\alpha}$};
    \end{tikzpicture} &= \frac{(d-2)\Gamma(\tfrac12d-1)}{4\pi^{d/2}}\frac{x_{12}^{\;\;\; \mu}\sigma_{\mu\alpha\dot{\alpha}}}{(x_{12}^{\;\;\; 2})^{d/2}}\,,\\
    \begin{tikzpicture}[baseline=(vert_cent.base)]
        \node (vert_cent) at (0,0) {$\phantom{\cdot}$};
        \draw[dashed,shorten <=0.75pt] (0,0)--(0,0.75);
        \draw[solid]  (-0.6,-0.6)--(0,0);
        \draw[solid]  (0.6,-0.6)--(0,0);
        \node[xshift=-3pt,yshift=-3pt] at (-0.6,-0.6) {$a$};
        \node[xshift=3pt,yshift=-3pt] at (0.6,-0.6) {$b$};
        \node[xshift=0pt,yshift=6pt] at (0,0.75) {$i$};
        \node [fill, shape=rectangle, minimum width=4pt, minimum height=4pt, inner sep=0pt, anchor=center] at (0,0) {};
        \node[xshift=7pt] at (0,0) {$x$};
    \end{tikzpicture} &= -\mu^{\varepsilon/2}y_{iab}\int d^{\lsp d}x\,.
\end{aligned}
\end{equation}
In order to solve the integrals that arise from the diagrams in Fig.\ \ref{BetaDefScalarFermionDiag}, we will make repeated use of the integrals
\begin{equation}
    \begin{aligned}
        \int d^{\lsp d}x_3\,\frac{1}{(x_{13}^{\;\;\; 2})^{\Delta_1/2}(x_{23}^{\;\;\; 2})^{\Delta_2/2}}&=\frac{\pi^{d/2}}{(x_{12}^{\;\;\; 2})^{\frac{\Delta_1+\Delta_2-d}{2}}}\frac{\Gamma\big(\frac{\Delta_1+\Delta_2-d}{2}\big)\Gamma\big(\frac{d-\Delta_1}{2}\big)\Gamma\big(\frac{d-\Delta_2}{2}\big)}{\Gamma\big(\frac{\Delta_1}{2}\big)\Gamma\big(\frac{\Delta_2}{2}\big)\Gamma\big(\frac{2d-\Delta_1-\Delta_2}{2}\big)}\,, \\
        \int d\tau'\frac{1}{(\mathbf{x}^2+(\tau-\tau')^2)^\Delta}&=\frac{\sqrt{\pi}}{|\mathbf{x}|^{1-2\Delta}}\frac{\Gamma(\Delta-\frac{1}{2})}{\Gamma(\Delta)}\,.
    \end{aligned}
\end{equation}
To begin, we notice that, following section \ref{bulkrelation}, there is no need to perform any integration for the diagrams
$$
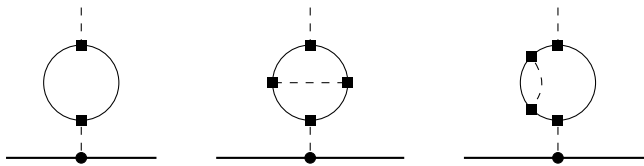
\begin{tikzpicture}[baseline=(vert_cent.base)]
        \node (vert_cent) at (0,1) {$\phantom{\cdot}$};
        \draw[dashed] (0,2)--(0,1.5);
        \draw[dashed,shorten <=2.5pt] (0,0.5)--(0,0);
        \draw[thick] (-1,0)--(1,0);
        \filldraw (0,0) circle (2pt);
        \node [fill, shape=rectangle, minimum width=4pt, minimum height=4pt, inner sep=0pt, anchor=center] at (0,1.5) {};
        \node [fill, shape=rectangle, minimum width=4pt, minimum height=4pt, inner sep=0pt, anchor=center] at (0,0.5) {};
        \draw (0,1) circle (0.5cm);
    \end{tikzpicture}\qquad
    \begin{tikzpicture}[baseline=(vert_cent.base)]
        \node (vert_cent) at (0,1) {$\phantom{\cdot}$};
        \draw[dashed] (0,2)--(0,1.5);
        \draw[dashed,shorten <=2.5pt] (0,0.5)--(0,0);
        \draw[thick] (-1.25,0)--(1.25,0);
        \filldraw (0,0) circle (2pt);
        \node [fill, shape=rectangle, minimum width=4pt, minimum height=4pt, inner sep=0pt, anchor=center] at (0,1.5) {};
        \node [fill, shape=rectangle, minimum width=4pt, minimum height=4pt, inner sep=0pt, anchor=center] at (0,0.5) {};
        \node [fill, shape=rectangle, minimum width=4pt, minimum height=4pt, inner sep=0pt, anchor=center] at (-0.5,1) {};
        \node [fill, shape=rectangle, minimum width=4pt, minimum height=4pt, inner sep=0pt, anchor=center] at (0.5,1) {};
        \draw[dashed,shorten <=0.5pt] (-0.5,1)--(0.5,1);
        \draw (0,1) circle (0.5cm);
    \end{tikzpicture}\qquad
    \begin{tikzpicture}[baseline=(vert_cent.base)]
        \node (vert_cent) at (0,1) {$\phantom{\cdot}$};
        \draw[dashed] (0,2)--(0,1.5);
        \draw[dashed,shorten <=2.5pt] (0,0.5)--(0,0);
        \draw[thick] (-1.25,0)--(1.25,0);
        \filldraw (0,0) circle (2pt);
        \node [fill, shape=rectangle, minimum width=4pt, minimum height=4pt, inner sep=0pt, anchor=center] at (0,1.5) {};
        \node [fill, shape=rectangle, minimum width=4pt, minimum height=4pt, inner sep=0pt, anchor=center] at (0,0.5) {};
        \draw[white] (0,1)--++(135:0.5) node (u) {};
        \draw[white] (0,1)--++(225:0.5) node (d) {};
        \node [fill, shape=rectangle, minimum width=4pt, minimum height=4pt, inner sep=0pt, anchor=center] at (u) {};
        \node [fill, shape=rectangle, minimum width=4pt, minimum height=4pt, inner sep=0pt, anchor=center] at (d) {};
        \draw[dashed,shorten <=-1.5pt] (u) to[out=-60,in=60] (d);
        \draw (0,1) circle (0.5cm);
    \end{tikzpicture}$$
and we can instead immediately write down the defect counterterms based on the bulk field renormalization:
\begin{equation}
    f_{1,i}^{(1)}\supset \frac{1}{16\pi^2}\frac{1}{2}Y_{ij}h_j-\frac{1}{(16\pi^2)^2}\Big(\frac{1}{4}\widetilde{Y}_{ijkj}h_k-\frac{3}{8}\widetilde{Y}_{ijjk}h_k\Big)\,,\qquad f_{1,i}^{(2)}\supset \frac{1}{(16\pi^2)^2}\big(2\lsp\widetilde{Y}_{ijkj}h_k+\widetilde{Y}_{ijjk}h_k\big)\,.
\end{equation}

Next, we must require
\begin{equation}\label{compldiag}
    \begin{tikzpicture}[baseline=(vert_cent.base)]
        \node (vert_cent) at (0,1) {$\phantom{\cdot}$};
        \draw[dashed] (0,2)--(0,1.5);
        \draw[dashed,shorten <=2.5pt] (0,0.5)--(0,0);
        \draw[thick] (-1.25,0)--(1.25,0);
        \filldraw (0,0) circle (2pt);
        \filldraw (-0.5,0) circle (2pt);
        \filldraw (0.5,0) circle (2pt);
        \node [fill, shape=rectangle, minimum width=4pt, minimum height=4pt, inner sep=0pt, anchor=center] at (0,1.5) {};
        \node [fill, shape=rectangle, minimum width=4pt, minimum height=4pt, inner sep=0pt, anchor=center] at (0,0.5) {};
        \draw[white] (0,1)--++(320:0.5) node (r) {};
        \draw[white] (0,1)--++(220:0.5) node (l) {};
        \node [fill, shape=rectangle, minimum width=4pt, minimum height=4pt, inner sep=0pt, anchor=center] at (l) {};
        \node [fill, shape=rectangle, minimum width=4pt, minimum height=4pt, inner sep=0pt, anchor=center] at (r) {};
        \draw (0,1) circle (0.5cm);
        \draw[dashed] (l)--(-0.5,0);
        \draw[dashed] (r)--(0.5,0);
    \end{tikzpicture}+
    \begin{tikzpicture}[baseline=(vert_cent.base)]
        \node (vert_cent) at (0,1) {$\phantom{\cdot}$};
        \draw[dashed] (0,2)--(0,0);
        \draw[thick] (-1.25,0)--(1.25,0);
        \filldraw (0,0) circle (2pt);
        \draw[dashed] (0,1)--(-0.625,0);
        \filldraw (-0.625,0) circle (2pt);
        \draw[dashed] (0,1)--(0.625,0);
        \filldraw (0.625,0) circle (2pt);
        \node[draw,shape=rectangle,minimum width=8.5pt,minimum height=8.5pt,black,cross,fill=white] at (0,1) {};
    \end{tikzpicture}+
    \begin{tikzpicture}[baseline=(vert_cent.base)]
        \node (vert_cent) at (0,1) {$\phantom{\cdot}$};
        \draw[dashed] (0,2)--(0,0);
        \draw[thick] (-1,0)--(1,0);
        \node[draw,circle,thick,black,cross,fill=white] at (0,0) {};
    \end{tikzpicture}=\;\text{finite in the limit }\varepsilon\to0\,,
\end{equation}
where now the bulk vertex counterterm is given by~\cite{Machacek:1984zw}
\begin{equation}
    \begin{tikzpicture}[baseline=(vert_cent.base)]
        \node (vert_cent) at (0,0) {$\phantom{\cdot}$};
        \draw[dashed,shorten <=0.75pt] (-0.5,-0.5)--(0.5,0.5);
        \draw[dashed,shorten <=0.75pt]  (-0.5,0.5)--(0.5,-0.5);
        \node[xshift=-3pt,yshift=3pt] at (-0.5,0.5) {$i$};
        \node[xshift=3pt,yshift=3pt] at (0.5,0.5) {$j$};
        \node[xshift=3pt,yshift=-3pt] at (0.5,-0.5) {$k$};
        \node[xshift=-3pt,yshift=-3pt] at (-0.5,-0.5) {$l$};
        \node[draw,shape=rectangle,minimum width=8.5pt,minimum height=8.5pt,black,cross,fill=white] at (0,0) {};
        \node[xshift=10pt] at (0,0) {};
    \end{tikzpicture}=-\frac{1}{16\pi^2}\frac{4}{\varepsilon}(\widetilde{Y}_{ijkl}+\widetilde{Y}_{ikjl}+\widetilde{Y}_{ijlk})\,.
\end{equation}
To calculate the poles of the first diagram in \eqref{compldiag} we follow \cite[Appendix B]{Giombi:2022vnz} and work in momentum space rather than coordinate space. The fermionic loop divides the integral into three different terms. It is important to note that while $\delta$ of \cite{Giombi:2022vnz} is not connected to the dimension $d$, we must take it to be equal to $-\varepsilon$. Thus, while they may safely drop terms linear in $\delta$ from the integrals, in our case such terms may interact with potential $\text{O}(\varepsilon^{-2})$ terms to affect the first order pole. The first integral is given by (B.6) in \cite{Giombi:2022vnz}, where it is evaluated in (B.9). Happily, this is already only $\text{O}(\varepsilon^{-1})$, so that we can simply borrow their result, which in our case gives a contribution
 \begin{equation}
     -\frac{1}{256\pi^4}\frac{\pi^2}{18\varepsilon}\widetilde{ Y}_{ijkl}h_jh_kh_l
 \end{equation}
to the counterterm. The other two terms we must compute explicitly. They are in fact both equal and the relevant integral to compute is
\begin{equation}
\begin{aligned}
    \tfrac12\mu^{3\varepsilon}\widetilde{Y}_{ijkl}h_jh_kh_l&\int\frac{d^dk}{(2\pi)^d}\frac{d^{d-1}\mathbf{k}_1}{(2\pi)^{d-1}}\frac{d^{d-1}\mathbf{k}_2}{(2\pi)^{d-1}}\frac{1}{\mathbf{k}_1^2\mathbf{k}_2^2(\mathbf{k}_1+\mathbf{k_2})^2(k-k_1)^2(k+k_2)^2}=\\
    &=\frac{1}{256\pi^4}\frac13\bigg(\frac{1}{\varepsilon^2}-\frac{1}{2\varepsilon}\bigg(2+3\gamma-3\log\Big(\frac{16\pi\mu^2}{m^2}\Big)\bigg)+\text{O}(\varepsilon^0)\bigg)\widetilde{Y}_{ijkl}h_jh_kh_l\,,
\end{aligned}
\end{equation}
where here $k_1^0=k_2^0=0$ and $m$ is a mass scale introduced to regulate an IR divergence. The IR divergence in these integrals is a consequence of working in momentum rather than position space. The $m$-dependence cancels with a corresponding term needed to regulate an identical IR divergence in the momentum space expression for the bulk counterterm graph (middle diagram in \eqref{compldiag}). After a short calculation, one then finds the full counterterm
\begin{equation}
   f_{3,i}^{(1)}\supset\frac{1}{(16\pi^2)^2}\frac13\Big(1-\frac16\pi^2\Big)\widetilde{Y}_{ijkl}h_jh_kh_l\,, \qquad f_{3,i}^{(2)}\supset-\frac{1}{(16\pi^2)^2}\frac{1}{3}\widetilde{Y}_{ijkl}h_{j}h_kh_l\,.
\end{equation}
Note that $\frac16\pi^2$ is equal to $\zeta_2$ or $\text{Li}_2(1)$. The $1/\varepsilon^{2}$ term agrees with the 't Hooft relations~\cite{tHooft:1973mfk}, whose general form is here obtained from \eqref{bareindep} at order $1/\varepsilon^n$ for $n\geq 1$:
\begin{equation}
\begin{aligned}
        -\bigg(\frac{1}{2}\Big(1-h_j\frac{\partial}{\partial h_j}-y_{jab}\frac{\partial}{\partial y_{jab}}&-y^*_{jab}\frac{\partial}{\partial y^*_{jab}}\Big)-\lambda_{jklm}\frac{\partial}{\partial\lambda_{jklm}}\bigg)\sum_p f_{p,i}^{(n+1)}\\
        &=\Big(\hat{\beta}_j\frac{\partial}{\partial h_j}+\hat{\beta}_{jab}\frac{\partial}{\partial y_{jab}}+\hat{\beta}^*_{jab}\frac{\partial}{\partial y^*_{jab}}+\hat{\beta}_{jklm}\frac{\partial}{\partial\lambda_{jklm}}\Big)\sum_p f_{p,i}^{(n)}\,,
    \end{aligned}
\end{equation}
where $\hat{\beta}$ are the standard quantum corrections to the beta functions; see e.g.\ \cite{Jack:1983sk, Jack:1984vj, Machacek:1983fi, Machacek:1984zw, Jack:2023zjt}. One then sees that for this diagram only the $\widetilde{Y}_{ijkl}$ term from the quartic coupling beta function will contribute on the right-hand side of the 't Hooft relations.

Finally, we must demand
\begin{equation}\label{ferdren4}
    \begin{tikzpicture}[baseline=(vert_cent.base)]
        \node (vert_cent) at (0,1) {$\phantom{\cdot}$};
        \draw[dashed] (0,2)--(0,0);
        \draw[thick] (-1.25,0)--(1.25,0);
        \filldraw (0,0) circle (2pt);
        \draw[white,name path=LP] (0,1.4)--(-1,0) node[midway] (l) {};
        \node [fill, shape=rectangle, minimum width=4pt, minimum height=4pt, inner sep=0pt, anchor=center] at (0,1.4) {};
        \draw[dashed,shorten <=-2pt] (0,1.4)--(1,0);
        \filldraw (1,0) circle (2pt);
        \filldraw (-1,0) circle (2pt);
        \draw[fill=white,name path=CIR] (l) circle (0.3cm);
        \node [name intersections={of=LP and CIR}, fill, shape=rectangle, rotate=52, minimum width=4pt, minimum height=4pt, inner sep=0pt, anchor=center] at (intersection-1) {};
        \node [name intersections={of=LP and CIR}, fill, shape=rectangle, rotate=52, minimum width=4pt, minimum height=4pt, inner sep=0pt, anchor=center] at (intersection-2) {};
        \draw[dashed,name intersections={of=LP and CIR},shorten <=-2pt] (0,1.4)--(intersection-1);
        \draw[dashed,name intersections={of=LP and CIR},shorten <=-2pt] (intersection-2)--(-1,0);
    \end{tikzpicture}+
    \begin{tikzpicture}[baseline=(vert_cent.base)]
        \node (vert_cent) at (0,1) {$\phantom{\cdot}$};
        \draw[thick] (-1.25,0)--(1.25,0);
        \node [fill, shape=rectangle, minimum width=4pt, minimum height=4pt, inner sep=0pt, anchor=center] at (0,1.4) {};
        \draw[dashed] (0,2)--(0,0);
        \draw[dashed,shorten <=-2pt] (0,1.4)--(-1,0);
        \draw[dashed,shorten <=-2pt] (0,1.4)--(1,0);
        \node[draw,shape=rectangle,minimum width=8.5pt,minimum height=8.5pt,black,cross,fill=white] at (-0.5,0.7) {};
        \filldraw (0,0) circle (2pt);
        \filldraw (1,0) circle (2pt);
        \filldraw (-1,0) circle (2pt);
    \end{tikzpicture}+
    \begin{tikzpicture}[baseline=(vert_cent.base)]
        \node (vert_cent) at (0,1) {$\phantom{\cdot}$};
        \draw[dashed] (0,2)--(0,0);
        \draw[thick] (-1,0)--(1,0);
        \node[draw,circle,thick,black,cross,fill=white] at (0,0) {};
    \end{tikzpicture}=\;\text{finite in the limit }\varepsilon\to0\,,
\end{equation}
where the middle graph in the left-hand side includes a bulk propagator correction to the $\text{O}(\lambda)$ diagram. Note that only the diagrams where the correction lies on an internal leg will contribute to the beta function, for placing it on the external leg would only lead to non-overlapping divergences that would be totally cancelled by already determined counterterms. The counterterm we use for the propagator correction is given from \eqref{Zphi}:
\begin{equation}
    \begin{tikzpicture}[baseline=(vert_cent.base)]
        \node (vert_cent) at (0,0) {$\phantom{\cdot}$};
        \draw[dashed] (0,0)--(2,0);
        \node[xshift=-4pt,yshift=0pt] at (0,0) {$i$};
        \node[xshift=4pt,yshift=0pt] at (2,0) {$j$};
        \node[draw,shape=rectangle,minimum width=8.5pt,minimum height=8.5pt,black,cross,fill=white] at (1,0) {};
    \end{tikzpicture}=-\frac{1}{16\pi^2}\frac{1}{\varepsilon}Y_{ij}\,.
\end{equation}
One then finds that \eqref{ferdren4} fixes
\begin{equation}
    f_{3,i}^{(1)}\supset-\frac{1}{(16\pi^2)^2}\frac{1}{12}\lambda_{ijkl}Y_{lm}h_jh_kh_m\,,\qquad f_{3,i}^{(2)}\supset\frac{1}{(16\pi^2)^2}\frac{1}{12}\lambda_{ijkl}Y_{lm}h_jh_kh_m\,.
\end{equation}
Combining these defect counterterms, one finds the beta function given in Eq.\ \eqref{BetaDefScalarFermion} in the text with the use of the extension of Eq.~\eqref{betafromZ} to include Yukawa couplings, namely
\begin{equation}
    \hat{\beta}_i=-\bigg(\frac12\Big(1-h_j\frac{\partial}{\partial h_j}-y_{jab}\frac{\partial}{\partial y_{jab}}-y^*_{jab}\frac{\partial}{\partial y^*_{jab}}\Big)-\lambda_{jklm}\frac{\partial}{\partial\lambda_{jklm}}\bigg)\sum_p f_{p,i}^{(1)}\,,
\end{equation}
and after rescaling $y\to 4\pi\lsp y$, $\lambda\to 16\pi^2\lsp\lambda$.

\end{appendices}

\bibliography{main}

\end{document}